\documentclass[aps,twocolumn,showpacs,preprintnumbers,amsmath,amssymb]{revtex4}
\usepackage{latexsym}
\usepackage{psfig,graphicx}
\usepackage{bm}

\renewcommand{\theequation}{\arabic{section}.\arabic{equation}}
\newcommand{\bea}{\begin{eqnarray}}
\newcommand{\eea}{\end{eqnarray}}
\newcommand{\be}{\begin{equation}}
\newcommand{\ee}{\end{equation}}
\newcommand{\pkt}{\; .}

\newcommand{\eqn}[1]{(\ref{#1})}

\newcommand{\tr}{{\rm Tr}}

\newcommand{\calm}{{\cal M}}

\newcommand{\cals}{{\cal S}}

\newcommand{\call}{{\cal L}}

\newcommand{\bfx}{{\bf x}}

\newcommand*{\del}{\partial}

\begin{document}
\preprint{DO-TH-02/22}
\date{December 20, 2002; revised February 2002}
\title{\bf Nonequilibrium evolution of $\Phi^4$ theory in $1+1$ dimensions
in the $2PPI$ formalism}
\author{J\"urgen Baacke}
\email{baacke@physik.uni-dortmund.de}
\author{Andreas Heinen}
\email{andreas.heinen@uni-dortmund.de}
\affiliation{Institut f\"ur Physik, Universit\"at Dortmund,
D - 44221 Dortmund , Germany}
\begin{abstract}
\noindent
We consider the out-of-equilibrium evolution of a classical condensate 
field and its quantum fluctuations for a $\Phi^4$ model in $1+1$ dimensions
 with a symmetric and a double well potential. 
We use the 2PPI formalism and go beyond the Hartree approximation
by including the sunset term. In addition 
to the mean field $\phi(t)=\left<\Phi\right>$ the 2PPI formalism 
uses as variational
parameter a time dependent mass $\calm^2(t)$ which contains all local
insertions into the Green function. We compare our results to those obtained
in the Hartree approximation. In the symmetric $\Phi^4$ theory we observe
that the mean field shows a stronger dissipation than the one
found in the Hartree approximation.
The dissipation is roughly exponential in an intermediate time region.
In the theory with spontaneous symmetry breaking, i.e., with a double well
potential, the field amplitude tends to zero, i.e., to the symmetric
configuration. This is expected on general grounds: in $1+1$ dimensional 
quantum field theory there is  no spontaneous symmetry breaking
for $T >0$, and so there should be none at finite energy density 
(microcanonical ensemble), either.
 Within the time range of our simulations the momentum spectra do not 
thermalize and display parametric resonance bands.
\end{abstract}
\pacs{03.65.Sq, 05.70.Fh, 11.30.Qc}
\maketitle

\section{\label{intro}Introduction}
\setcounter{equation}{0}

There has recently been considerable activity in investigating
the nonequilibrium evolution of quantum field theory beyond the large-$N$
approximation. In particular there has been formulated
\cite{Berges:2001fi,Aarts:2002dj} a systematic approach
(2PI-NLO) in which all next-to leading order contributions in the
$1/N$ expansion are included, using the CJT or 2PI formalism
\cite{Cornwall:1974vz} generalized to nonequilibrium evolution
in Ref. \cite{Calzetta:1987ey}, using the closed time path (CTP)
or Schwinger-Keldysh \cite{skform} formalism.
 A similar approximation including some but not all
contributions of next-to-next-to leading order is the 
bare vertex approximation (BVA) \cite{Mihaila:2001sr}. Numerical 
simulations have been 
performed mostly in $1+1$ dimensions, for the classical or quantum
 $\Phi^4$ theory with a symmetric \cite{Berges:2001fi,
Aarts:2001yn,Blagoev:2001ze,Cooper:2002ze,Berges:2000ur} and 
with a double well \cite{Cooper:2002qd} potential.
A first simulation in an O$(N)$ model for $N\neq 1$ in $3+1$ dimensions
has just appeared \cite{Berges:2002cz}, for a symmetric potential. 
The analysis of models with  $N\neq 1$ and with spontaneous symmetry
breaking, i.e., a Mexican hat potential, should be very important for 
understanding the r\^ole of the Goldstone modes and their
influence on the phase structure of the theory in a certain
approximation.

A more modest step beyond leading order large $N$  has been 
taken in Ref. \cite{Baacke:2001zt},
where the Hartree approximation was used in an O$(N)$ model in
$3+1$ dimensions. This approach includes some terms of the nonleading
orders, but not all of them. Our investigation made evident the role 
of parametric resonance in the system of ``sigma'' and
``pion'' modes, and the role of the Goldstone modes in stabilizing 
the evolution in the regions where the equilibrium effective potential 
is complex, features still to be investigated in the systematic 
$1/N$ motivated approximations.

An approach for including nonleading orders 
in a self-consistent resummation scheme has been proposed
some time ago by Verschelde and collaborators
\cite{Verschelde:bs,Coppens:zc}, the so-called 2PPI
expansion. Like in the 2PI or CJT formalism the mean field and the
internal Green functions are determined self-consistently. As in the
2PI formalism the equations of motion follow from an effective action,
which here is a functional of a mean field $\phi=\left<\Phi\right>$ and
an effective mass $\calm$. In contrast
to the 2PI approach only local insertions into the Green function are 
resummed, so the Green function is, in all orders, a functional of 
the local mass term $\calm$ that in general will depend on $x=(t,\bfx)$.
In particular this Green function is different from the physical Green 
function. As far as the resummation is concerned the approach is less powerful:
one has to include more diagrams if one wants to reach the same order
in a loop or $1/N$ expansion as in the 2PI expansion. The fact that
the propagators have a simpler structure may be a disadvantage as by 
this structure the approximation is less flexible than 2PI. On the other hand
the calculations are technically less involved, in particular the
formalism does not require ladder resummations which complicate
the renormalization of the 2PI approach \cite{vanHees:2001ik}.    
In one-loop order the approach is equivalent to the Hartree approximation.
Recent progress in the 2PPI formalism includes the demonstration
or renormalizability \cite{Verschelde:2000ta,Verschelde:2000dz} and 
some finite-temperature
two-loop calculations in $3+1$ dimensions.
An interesting result was that for $N=1$ \cite{Smet:2001un} and
for $N\neq 1$ \cite{Baacke:2002pi} the order of the phase transition 
between the 
spontaneously broken and symmetric phases becomes second order
in the 2-loop approximation, while it is first order
in the Hartree approximation. The results for the 2PPI expansion have
been compared to exact results in \cite{Okopinska:1995mi} for the anharmonic oszillator; even more recently 
\cite{Dudal:2002zn} the two-loop
approximation has been compared, in $1+1$ dimensions, with exact
results of the Gross-Neveu model.

In this paper we will present the formulation and some numerical results for
the nonequilibrium evolution of the mean field and the self-consistent mass
in the 2PPI scheme, applied to $\Phi^4$ theory in $1+1$ dimensions.
We go beyond the one-loop (Hartree) approximation by
including the sunset graph, which represents the full two-loop contribution
in this formalism. We explicitly formulate a conserved energy functional 
which is used to monitor the numerical accuracy. As this is the first
investigation of the 2PPI formalism at two loops out of equilibrium we do
not attempt a detailed study; rather we aim at presenting the main new 
features of this approach, as well as compared to the one-loop Hartree 
approximation and as compared to the other  approaches
(2PI-NLO and BVA) mentioned above. We consider both the case of the
symmetric $\Phi^4$ potential and the case of the double well
potential which displays spontaneous symmetry breaking on the
classical as well as on the one-loop level.

The plan of the paper is as follows: In section 2 we formulate the model
and present the 2PPI formalism as applied to the system out of 
equilibrium. In section 3 we specify the two-loop approximation by  
giving the explicit expressions
for the basic graphs, the equations of motion, the conserved energy
and by discussing the initial conditions and renormalization.
In section 4 we give details of the numerical implementation. 
The numerical results are presented and discussed in section 5.
We end with conclusions and an outlook in section 6. The paper is
completed by two appendices.


\section{\label{formul}Formulation of the model}
\setcounter{equation}{0}
We consider the $\Phi^4$ quantum field theory defined by the Lagrange density
\be
  \call=\frac{1}{2} \del_\mu {\Phi}\,
  \del^\mu {\Phi} - \frac{1}{2}m^2\Phi^2-
  \frac{\lambda}{24}\Phi^4 
\pkt\ee
If $m^2> 0$  we refer to as the symmetric theory, and
with $m^2<0$ which we refer to as theory with spontaneous symmetry breaking.
These terms relate to the classical theory and do not imply the occurrence
of spontaneous symmetry breaking in the quantum field theory.

The 2PPI formalism proposed by Verschelde and Coppens 
\cite{Verschelde:bs,Coppens:zc} 
is based on an effective action which is formulated in terms
of a mean field $\phi$ and a local insertion $\Delta$. It is the Legendre
transformation of a generating functional with a source $J(x)$ for the 
field $\phi(x)$ and another {\em local} source $K(x)$ for $\phi^2(x)$. Here 
lies the difference to the well-known CJT formalism, where one introduces 
a bilocal source $K(x,x')$ for a Green function $G(x,x')$.
Graphically the 2PPI scheme resums all local insertion into a Green function
which in all orders remains a generalized free particle Green function,
or Green function in an external field, i.e.
\be \label{greendef}
 G^{-1}(x,x')=i \left[\Box + \calm^2(x)\right]\delta(x-x')\ ,
\ee
where
\be \label{calmdef}
\calm^2(x)=m^2+\frac{\lambda}{2}\phi^2(x) +\frac{\lambda}{2}
\Delta(x)
\pkt\ee
So in contrast to the 2PI formalism this Green function is not a 
variational object, it is a functional of $\calm^2(x)$. It is not 
the physical Green function. 

\begin{figure}
\begin{center}
(a)\includegraphics[scale=0.5]{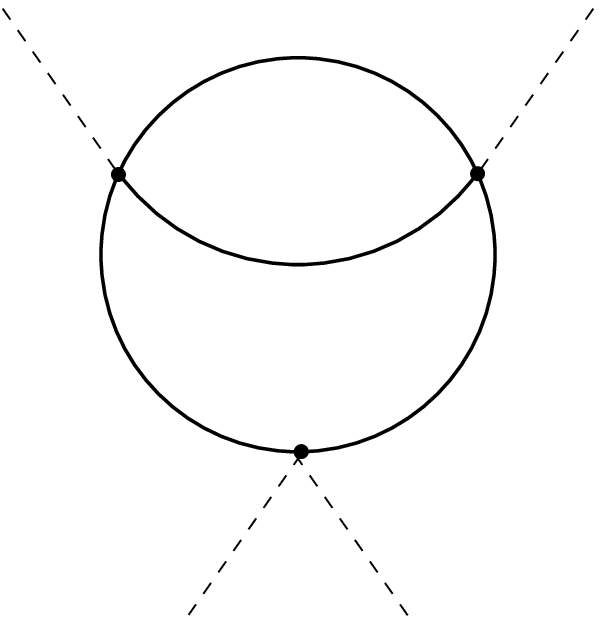}
\hspace{1cm}(b)\quad\includegraphics[scale=0.5]{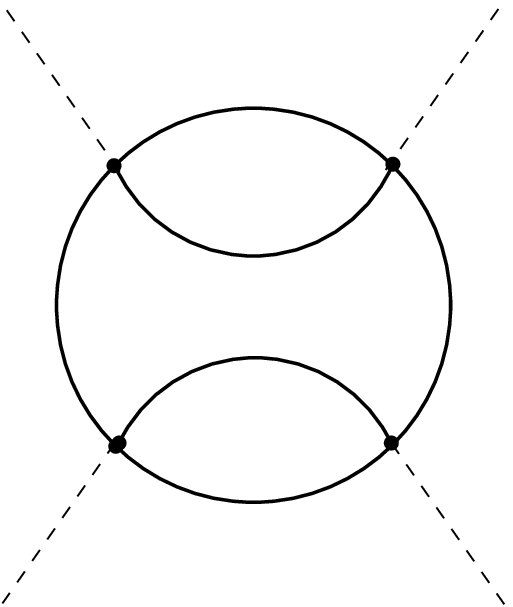}
\end{center}
\caption{\label{fig:2PPR2PR}
Examples for two particle point reducible (2PPR) and
  irreducible (2PPI)
  diagrams: (a) a diagram which is 2PR and 2PPR (b) a diagram which is
  2PR but 2PPI; solid lines: internal
propagators; dashed lines: external fields $\phi$.}
\end{figure}

The 2PPI formalism in its original form is based on the action
\be \label{gammaphidelta}
\Gamma[\phi,\Delta]=S_{\rm class}[\phi] +
\Gamma^{\rm 2PPI}[\phi,\calm^2] +\frac{\lambda}{8}\int d^2x \Delta^2(x)
\pkt\ee
Here $\Gamma^{\rm 2PPI}[\phi,\calm^2]$ is the sum of all 2PPI graphs;
these are defined as graphs which do not decay into two parts of 
two lines {\em joining at a point} are cut.  In the 2PI formalism
one includes into the analogous $\Gamma^{\rm 2PI}$ all graphs which 
do not decay into two parts if {\em any} two lines are cut.
In order to visualize the difference  we show 
examples of 2PR and 2PPR graphs in Fig.~\ref{fig:2PPR2PR}.
It should be emphasized that in the 2PPI formalism the lines in the
graphs are Green functions with the variational mass term $\calm^2(x)$
as defined via Eq. \eqn{greendef}, while in the 2PI formalism the 
internal lines  refer to the variational Green functions 
induced by the {\em bilocal} sources.
 When comparing the sets of irreducible graphs
in both formalisms one has to take into account this difference
in the meaning of internal lines.

The insertion $\Delta(x)$ is given by
\be
\Delta(x)=-2 
\frac{\delta \Gamma^{\rm 2PPI}[\phi,\calm^2]}{\delta \calm^2(x)} \ .
\ee
As we have stated before it is simpler to formulate the action in terms 
of $\phi$ and $\calm^2$. We solve Eq. \eqn{calmdef} with respect to
$\Delta$ and insert this into Eq. \eqn{gammaphidelta}. Using the explicit
form of $S_{\rm class}$ we obtain
\bea \nonumber
\Gamma[\phi,\calm^2]&=&
\int d^2x \left[\frac{1}{2} \del_\mu \phi(x)\,
  \del^\mu \phi(x) - \frac{1}{2}\calm^2(x)\phi^2(x)\right.
\nonumber\\
&&\hspace{1cm}+\left.\frac{\lambda}{12}\phi^4(x)\right]
\nonumber\\
&& + \frac{1}{2\lambda}\int d^2 x \left[\calm^2(x)-m^2\right]^2
\nonumber\\
\label{gammaphicalm} 
&& + \Gamma^{\rm 2PPI}[\phi,\calm^2]
\pkt
\eea 
One easily checks that the equations of motion obtained by varying this action
with respect to $\phi(x)$ and $\calm^2(x)$ take the form
\bea
\label{eqphi}
0&=&\Box \phi + \calm^2(x)\phi(x) -\frac{\lambda}{3}
\phi^3(x)\nonumber \\ 
&&\hspace{2cm}-\frac{\delta \Gamma^{\rm 2PPI}[\phi,\calm^2]}{\delta \phi(x)}
\\
\label{eqcalm2}
\calm^2(x)&=&m^2+\frac{\lambda}{2}\phi^2(x)
-\lambda\frac{\delta \Gamma^{\rm 2PPI}[\phi,\calm^2]}{\delta \calm^2(x)}
\pkt\eea
The latter equation being identical to Eq. \eqn{calmdef}, we will refer
to it as gap equation.

\begin{figure}
\begin{center}
\includegraphics[width=7cm]{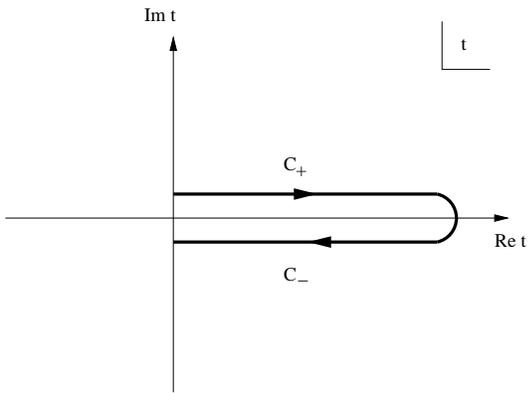}
\end{center}
\caption{\label{fig:CTP}
The closed time path in the complex $t$ plane.}
\end{figure}

Here we will consider states of the system that are spatially homogeneous;
so in Eqs. \eqn{eqphi} and \eqn{eqcalm2} the arguments $x=(t,x)$ should 
be replaced simply by $t$. Furthermore, in Eq. \eqn{gammaphicalm}
the space integration simply gives a volume (length) factor, and
in the nonequilibrium formalism the time integration
should be replaced by the closed time path (CTP) displayed in 
Fig.~\ref{fig:CTP}.

The equation for the Green function separates into space
and time dependence. Using the homogeneity of the state
in  space the Green function can be written as
\be
G(t,t';x,x')=\int_{-\infty}^{\infty} \frac{dp}{2\pi}
e^{ip(x-x')}G(t,t';p) \ .
\ee
As the equation is local in time the
Green function $G(t,t';p)$ can be expressed in terms of mode functions
\bea
G(t,t';p)&=&\frac{1}{2\omega_p}\left[
f(t,p)f^*(t',p)\Theta(t-t')\right.\nonumber\\
&&\qquad \left.+f(t',p)f^*(t,p)\Theta(t'-t)\right]
\eea
where $\omega_p$ is defined below, and where
the mode functions $f(t,p)$ satisfy 
\be \label{modeq}
\ddot f(t,p)+\left[p^2+\calm^2(t)\right] f(t,p)=0
\pkt
\ee
Here we choose the initial conditions for the mode functions at $t=0$ 
as for wave functions of free particles with mass
\be
m_0^2=\calm^2(0)=m^2+\frac{\lambda}{2}\phi^2(0)+\frac{\lambda}{2}\Delta(0)
\pkt\ee
This means that $f(0,p)=1$ and $\dot f(0,p)=-i\omega_p$
with $\omega_p=\sqrt{p^2+m_0^2}$.
This defines an initial Fock space. We continue the discussion 
on initial conditions in section \ref{subsec:renini}. 

In the CTP formalism \cite{skform} one uses Green functions with different time
orderings. The Green function $G(t,t';p)$ as defined above is
identical to  the Green function $G_>(t,t';p)$ with normal time
ordering; the anti-time-ordered Green function $G_<(t,t';p)$ 
is given by $G_<(t,t';p)=G_>(t',t;p)$. In the explicit formulae
one can use this identity in order to express  all relevant Green 
functions  by $G(t,t',p)$.


\section{\label{approx}
One-loop and two-loop contributions}
\begin{figure}
\begin{center}
(a)\includegraphics[scale=0.5]{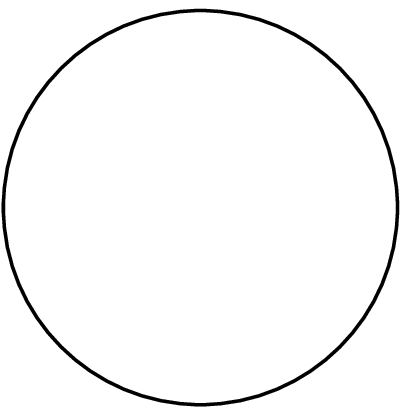}
\hspace{1cm}(b)\includegraphics[scale=0.5]{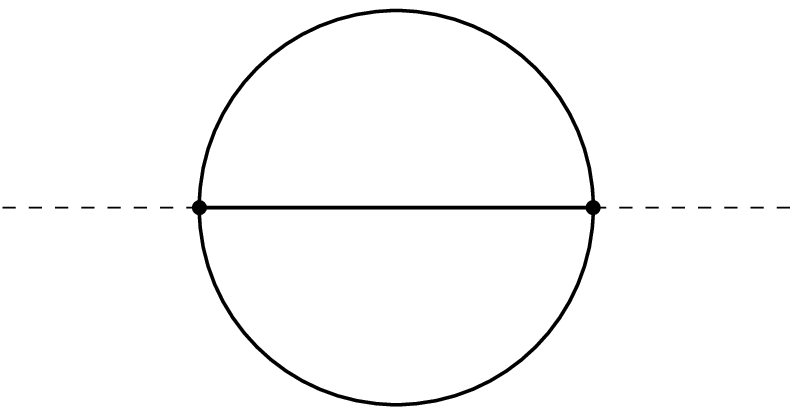}
\end{center}
\caption{\label{fig:vacgraphs}Bubble and sunset diagrams. 
We display the leading diagrams in 
the 2PPI action $\Gamma^{{\rm 2PPI}}$: (a) the bubble diagram;
(b) the sunset diagram; solid lines: internal
propagators; dashed lines: external fields $\phi$.}
\end{figure}

\setcounter{equation}{0}
Having established the general formalism we can now discuss the 
leading terms in a loop expansion. The two relevant graphs
are depicted in Fig.~\ref{fig:vacgraphs}, graphs a and b. The leading 
bubble diagram is the ``log det'' contribution. It leads to the tadpole 
insertion into the Green function. The next diagram represents 
the only contribution 
on the two-loop level. We will separately discuss these two contributions 
in the following.

\subsection{The bubble diagram}
The bubble diagram defines the leading Hartree contribution.
It is independent of $\phi$ and so does not yield an explicit
contribution to the equation of motion for $\phi$. Of course it enters,
indirectly, via its contribution to $\calm^2$.
Its contribution to $\Gamma^{\rm 2PPI}(\phi,\calm^2)$ is given by
\be
\Gamma^{(1)}[\phi,\calm^2]=\frac{i}{2}\tr\ln
\left[G^{-1}[\calm^2]\right] \ .
\ee
Its functional derivative with respect to $\calm^2$ is given by
\bea \label{delta1}
\frac{\delta \Gamma^{(1)}[\phi,\calm^2]}{\delta \calm^2(t)}
&=&-\frac{1}{2}\Delta^{(1)}(t)=-\frac{1}{2} \int\frac{dp}{2\pi}G(t,t;p)\nonumber\\
&=&
-\frac{1}{2}\int\frac{dp}{2\pi2\omega_p}\left|f(t,p)\right|^2 \ .
\eea
The energy density can be derived \cite{ItzyksonZuber}
by considering a variation
of the action under  $t \to t +\tau(t)$, 
which induces $\delta\phi(t)=\dot\phi(t)\tau(t)$,
$\delta \dot \phi(t)=\ddot\phi(t)\tau(t)+\dot\phi(t)\dot\tau(t)$
and $\delta \calm^2(t)=\tau(t)d\calm^2(t)/dt$.
The one-loop action only depends on $\calm^2(t)$. 
One then finds that the contribution of the bubble graph to the energy 
is defined by the relation
\bea \label{E1def}
\frac{dE^{(1)}(t)}{dt}&=& - \frac{\delta \Gamma^{(1)}[\phi,\calm^2]}
{\delta \calm^2(t)}\frac{d\calm^2}{dt}\nonumber\\
&=&
\frac{1}{2}\int \frac{dp}{2\pi}
G(t,t;p)\frac{d\calm^2}{dt}
\pkt
\eea
This equation can be integrated explicitly;
indeed one checks easily, using the mode equation
\eqn{modeq}, that the naive quantum energy defined by
\bea \label{E1}
E^{(1)}(t)&=&\frac{1}{2}\int \frac{dp}{2\pi 2\omega_p}
\bigg\{\left|\dot f(t,p)\right|^2 \nonumber \\
&&\qquad
+\left[p^2+\calm^2(t)\right]\left|f(t,p)\right|^2 \bigg\}
\pkt
\eea
is consistent with the defining equation  \eqn{E1def}.
This is of course well-known.
If only this one-loop contribution is included, 
the approximation is referred to as Hartree approximation.


\subsection{\label{subsec:sunset}The sunset diagram}

\begin{figure}
\begin{center}
(a)\includegraphics[scale=0.5]{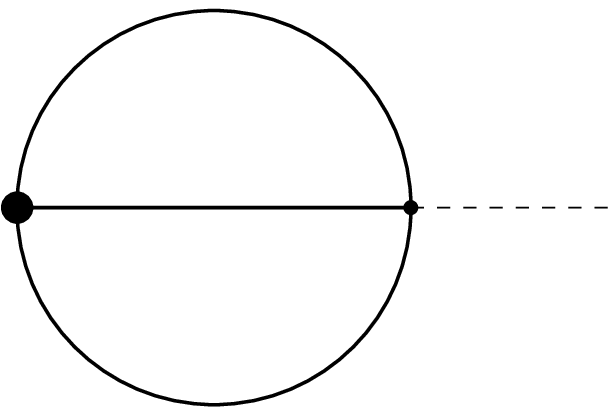}
\hspace{1cm}(b)\includegraphics[scale=0.5]{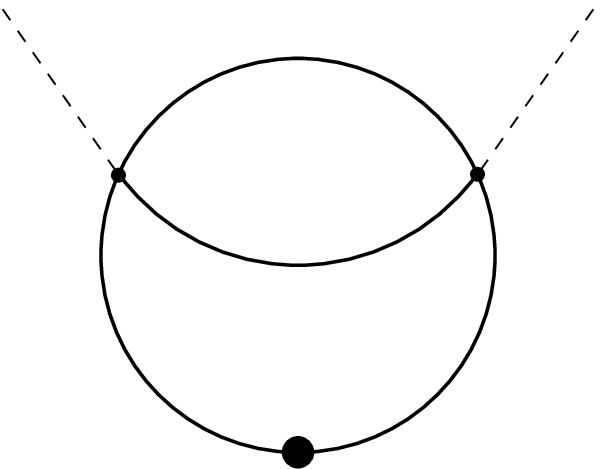}
\end{center}
\caption{\label{fig:calscalm}
Two-loop graphs in the equations of motion. 
(a) the amputated sunset diagram which appears in the equation of motion
\eqn{eqphi}
for the mean field $\phi$; 
(b) the tadpole diagram with fish insertion which
contributes to the gap equation \eqn{eqcalm2}; 
in both diagrams
the solid dots indicate the external time, the time variables of the other
vertices appear in the internal integrations; solid lines: internal
propagators; dashed lines: external fields $\phi$.}
\end{figure}

The unique two-loop contribution to $\Gamma^{\rm 2PPI}$ is 
the sunset diagram which in the CTP formalism is explicitly given by
\begin{widetext}
\be
\Gamma^{(2)}[\phi,\calm^2] =i \frac{\lambda^2}{12}
\int dx\, dx'
\int dt\, \phi(t)\int dt'\, 
G_P^3(t,t';x,x')\phi(t')
\ee
where the $t$ and $t'$ integrations are over a CTP contour and where 
$G_P$ is the path-ordered Green function.

The functional derivative with respect to $\phi(t)$ is given by
\be
\frac{\delta \Gamma^{(2)}[\phi,\calm^2]}{\delta \phi(t)}=-\cals(t)
\ee
with
\be
\cals(t)=
-i\frac{\lambda^2}{6}
\int_0^t dt'\phi(t')\int\prod_{\ell=1}^3 \left(\frac{dp_\ell}{2\pi}\right)
2\pi \delta\left(\sum_{\ell=1}^3 p_\ell\right)\left[\prod_{\ell=1}^3 G(t,t';p_\ell)
-\prod_{\ell=1}^3 G(t',t;p_\ell)\right] \label{eq:cals}\ .
\ee
This contribution to the equation of motion for $\phi(t)$, an 
amputated sunset diagram is 
represented graphically in Fig.~\ref{fig:calscalm}a.

The functional derivative with respect to $\calm^2(t)$ is given by
\bea \nonumber
\frac{\delta \Gamma^{(2)}[\phi,\calm^2]}{\delta \calm^2(t)}
&=&-\frac{1}{2}\Delta^{(2)}(t)
\label{eq:Delta2}\\  
&=&\frac{\lambda^2}{2}
\int_0^{t}{dt'}\phi(t')\int_0^{t'}{dt''}\phi(t'')
\int\prod_{\ell=1}^3\left(\frac{dp_\ell}{2\pi}\right)
2\pi\delta\left(\sum_{\ell=1}^3p_\ell\right)\nonumber
  \\
  &&\times   \left[G(t,t';p_3)- G(t',t;p_3)\right] \\
&&\times
  \left[G(t',t'';p_1)G(t',t'';p_2)G(t,t'';p_3)-
G(t'',t';p_1)G(t'',t';p_2)G(t'',t;p_3)\right]\nonumber\pkt
\eea
\end{widetext}
This graph which contributes to the gap equation is depicted in Fig.
\ref{fig:calscalm}b. It is a tadpole diagram with fish insertion.

Considering again a variation $t \to t+\tau(t)$ one finds
the contribution of the sunset term to the energy to be defined by
\be \label{E2def}
\frac{dE^{(2)}(t)}{dt}=
\dot\phi(t)\cals(t)+\frac{1}{2}\frac{d\calm^2(t)}{dt}\Delta^{(2)}(t) \ .
\ee
As far as we see this expression cannot be integrated explicitly; but the
relation can be integrated numerically to obtain $E^{(2)}$.
\be \label{E2}
E^{(2)}(t)=\int_0^t dt'
\left[\dot\phi(t')\cals(t')+\frac{1}{2}\frac{d\calm^2(t')}{dt'}
\Delta^{(2)}(t')\right]
\ee

\subsection{Equations of motion and energy}

Having determined the functional derivatives of  the action with respect to
$\phi(t)$ and $\calm^2(t)$ we can explicitly write down the equations of motion in 
the two-loop approximation:
\bea
0&=&\ddot\phi(t)+\calm^2(t)\phi(t)-\frac{\lambda}{3}\phi^3(t)+\cals(t)
\\
\calm^2(t)&=&m^2+\frac{\lambda}{2}\left[\phi^2(t)+\Delta^{(1)}(t)+
\Delta^{(2)}(t)\right] \label{eq:M2cal}\ .
\eea
We define the ``classical'' energy as the zero-loop expression
\bea
E^{(0)}(t)&=&
\frac{1}{2} \dot\phi^2(t)+ \frac{1}{2}\calm^2(t)\phi^2(t)-
  \frac{\lambda}{12}\phi^4(t)\nonumber \\
&&\qquad-
\frac{1}{2\lambda} \left[\calm^2(t)-m^2\right]^2
\pkt
\eea
We have defined the one-loop and two-loop contributions in
Eqs. \eqn{E1} and \eqn{E2}. One can check, using the equations of motion,
that the total energy
\be \label{etot}
E_{\rm tot}=E^{(0)}+E^{(1)}+E^{(2)}
\ee
is conserved.


\subsection{Renormalization and initial conditions}
\label{subsec:renini}
There is a wide choice of initial conditions for the
system. So one may choose an initial mean field
$\phi(0)$ and one may modify the Green function
by including contributions from the kernel
of $G^{-1}(x,x')$. In the one-loop approximation
the latter possibility is 
equivalent to choosing initial ensembles for which the
modes $f(t,p)$ are populated, or to Bogoliubov rotations
of the initial Fock space. In the two-loop approximation
this simple particle picture is no longer appropriate.

 However the choice of initial conditions is not entirely arbitrary
because of initial singularities \cite{Cooper:1987pt,Maslov:1998bf,
Baacke:1998zz}.
Starting with some nonzero value of $\phi(0)$ and with $m_0$ a solution
of the gap equation no initial time singularities are encountered.
So such a choice is a physically acceptable one.
In order to solve the gap equation at $t=0$ we have to know 
the contributions
$\Delta^{(1)}(0)$ and $\Delta^{(2)}(0)$. $\Delta^{(2)}$
has already be defined such that it vanishes at $t=0$; by this
choice we erase the memory of the past.
$\Delta^{(1)}$ is given by an integral over the fluctuations, so
it does not vanish. Furthermore it is divergent and so we have to discuss
renormalization.

In  $\Phi^4$ theory in $1+1$ dimensions there is only one
primitive divergence, the one of the tadpole graph. Renormalization
reduces, therefore, to making a shift in the tadpole term which
can be absorbed by a shift in the mass. 
Using dimensional regularization we rewrite  
the tadpole contribution $\Delta^{(1)}$, see Eq.
\eqn{delta1}, as

\be
\Delta^{(1)}(t)=\Delta^{(1)}_{\rm fin} (t)+
\frac{1}{4\pi}\left\{\frac{2}{\epsilon}-\gamma
+\ln\frac{4\pi\mu^2}{m_0^2}
\right\}
\ee
with the finite part of $\Delta^{(1)}$ defined
as
\be
\Delta^{(1)}_{\rm fin} (t)
= \int\frac{dp}{2\pi2\omega_p}\left[\left|f(t,p)\right|^2 
-1\right] \ .
\ee
This is the expression used in the numerical computation.

Including a mass counter term the gap equation 
now takes the form
\bea
\calm^2(t)&=&m^2 + \delta m^2 \nonumber \\
&&+\frac{\lambda}{2}
\bigg[\phi^2(t)+\Delta^{(1)}_{\rm fin}(t) \\
&&\qquad+\frac{1}{4\pi}\left\{\frac{2}{\epsilon}-\gamma
+\ln\frac{4\pi\mu^2}{m_0^2}
\right\}+\Delta^{(2)}(t)\bigg]\nonumber \ .
\eea
Choosing
\be 
\delta m^2=-\frac{\lambda}{8\pi}\left\{\frac{2}{\epsilon}-\gamma
+\ln\frac{4\pi\mu^2}{m^2}\right\}
\ee
the finite gap equation takes the form
\be
\calm^2(t)=m^2 + \delta m^2_{\rm fin}+   \frac{\lambda}{2}
\left[\phi^2(t)+\Delta^{(1)}_{\rm fin}(t)
+\Delta^{(2)}(t)\right]
\ee
with
\be
\delta m^2_{\rm fin} = \frac{\lambda}{8 \pi}
\ln \frac{m^2}{m_0^2}
\pkt
\ee
Initial conditions and renormalization are equivalent
to those in the one-loop approximation, which facilitates
a comparison between the two-loop 2PPI and the Hartree approximation. 


\section{Numerical Implementation}
\setcounter{equation}{0}
In the 2PPI approximation the Green function factorizes and
so one can work with mode functions. This considerably facilitates
the numerical computation of the ``memory'' integrals introduced
by the sunset graph. In particular one has to store only functions 
of one time argument, of course still for all times and all
momenta. The storage requirements grow only
linearly with time, so the evolution can be followed for relatively
long times. Furthermore the differential equations are ordinary
differential equations that can be solved precisely using
a Runge-Kutta algorithm. This can be important if one has to trace
parametric resonance phenomena.
Of course if the approximation itself is poor these numerical
advantages are useless. Still, the possibility of doing the calculations
with good precision allows to study the quality of the 
approximation reliably, including its possible shortcomings.

The time integration was done in steps of $\Delta t= 0.001$ to $0.005$.
The Wronskians of the mode functions were constant with a 
relative precision of $10^{-8}$.
For the momentum cutoff, which is a cutoff of a convergent integral,
we have chosen $p_{\rm max}=20$. As one can see from the momentum
spectra, this is a rather generous choice. It should be mentioned
that momentum conservation  leads to momenta that can 
be  beyond the cutoff.
This is a problem that can hardly be avoided, and a relatively
large momentum range should make such ``losses'' tolerable.
A more serious problem is the momentum grid. We observe
parametric resonance 
\footnote{We here use the term parametric resonance in a colloquial way,
not in the strict sense of a solution of Mathieu or Lam\'e equations
\cite{Traschen:1990sw,Kofman:1994rk,Boyanovsky:1996sq,
Baacke:2001zt}. Obviously, see Fig. 7,  the oscillations of the mass term 
lead, in spite of variations in amplitude, to a resonance-like enhancement 
that closely resembles the one found for true parametric resonance.}
, and this leads to amplitudes that vary
strongly in time {\em and momentum}. In the typical large-$N$ studies
this fact has lead the various groups to choose much finer
grids with several thousand momenta. This is not possible here,
we think that the essential features of the low momentum region
with parametric resonance and/or exponential growth subsist
with a less refined grids. This concerns in particular the 
self-stabilization of the system in the classically unstable
regions.
We have chosen $\Delta p = 0.05$, i.e., a grid of $400$ equidistant
momenta. In principle such a grid can lead
\cite{Destri:1999hd} to ``lattice artefacts'', corresponding here 
to a lattice size  $L= 2\pi/\Delta p = 40\pi $ in inverse mass units. 
Indeed we do not observe any phenomena that suggest such artefacts.
The choice of $\Delta p$ is also discussed in Appendix B.


\section{\label{sec:discuss}Discussion of the results}
\setcounter{equation}{0}
We have performed several simulations for the case
of a symmetric $\Phi^4$ potential and for a double well
potential which classically leads to spontaneous
symmetry breaking. The initial configuration
has been, in all cases, a mean field $\phi$ different from
its classical expectation value, and a quantum ensemble
corresponding to the ground state of a Fock space characterized
by an initial mass $m_0=\calm(0)$. We have obtained results for 
the time evolution of the mean field $\phi(t)$, for the self-consistent
mass $\calm^2(t)$ and for the energy. The relative importance
of the two-loop contributions can be seen in their
 contribution to $\calm^2$. In all cases we have 
compared the evolution with the one obtained in the Hartree 
approximation.

\subsection{Results for the symmetric $\Phi^4$ potential}

\begin{figure*}[htbp]
\begin{center}
(a)\hspace{-0.5cm}\includegraphics[width=7.5cm,height=4.9cm]{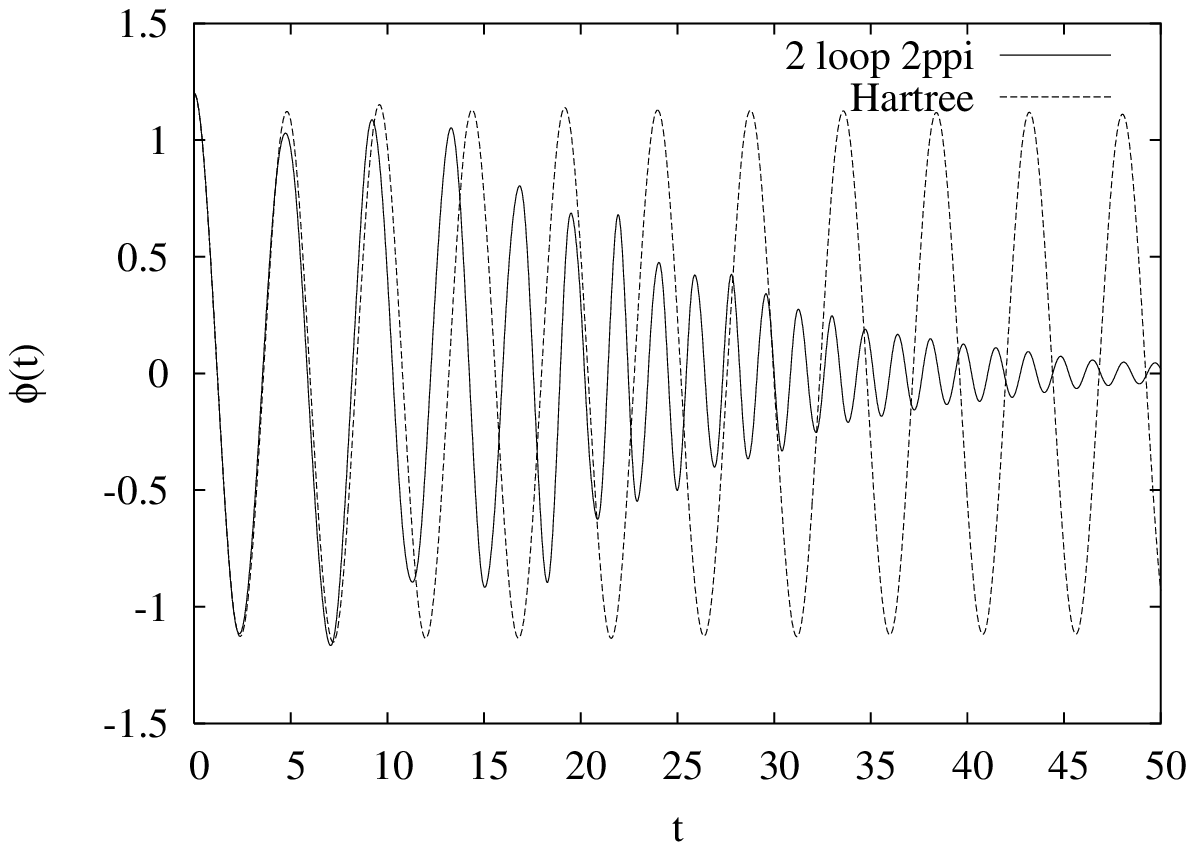}
\hspace{0.6cm}(b)\hspace{-0.5cm}\includegraphics[width=7.5cm,height=4.9cm]{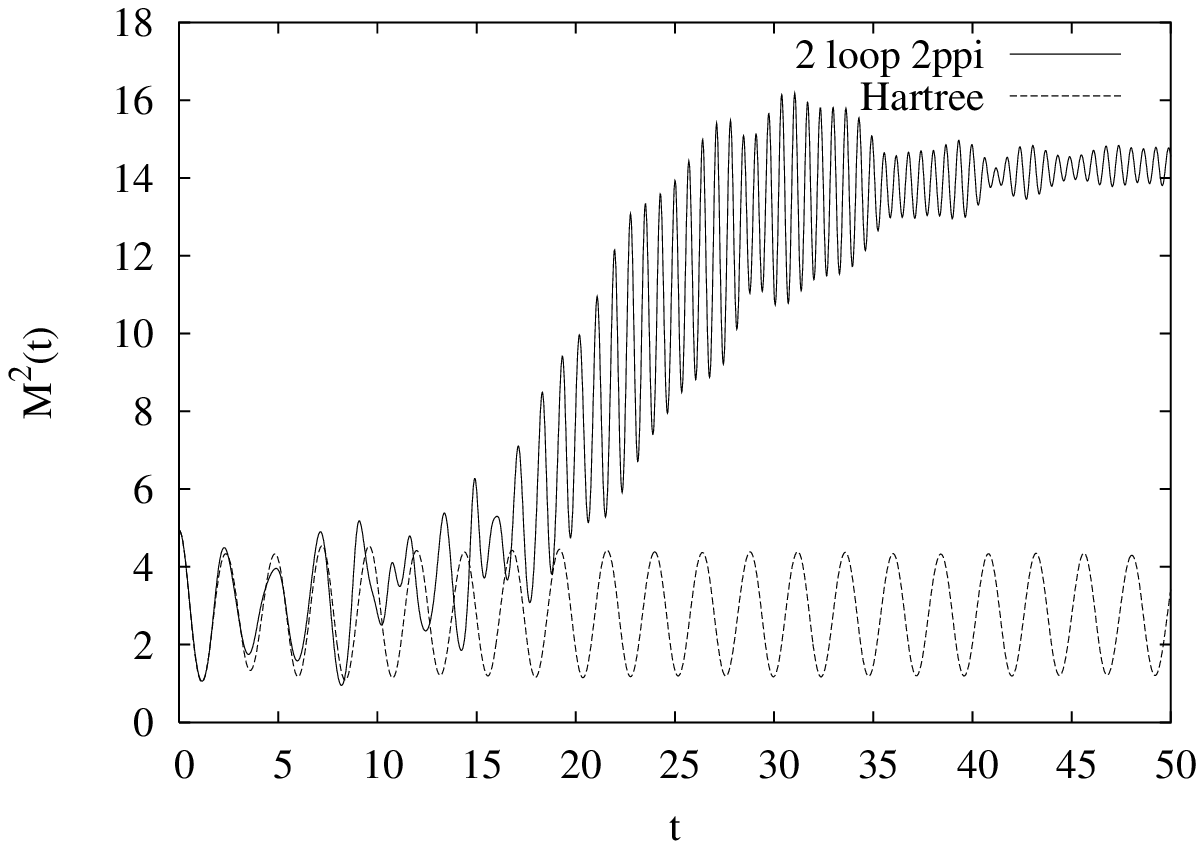}
\vspace{0.2cm}
\par
(c)\hspace{-0.5cm}\includegraphics[width=7.5cm,height=4.9cm]{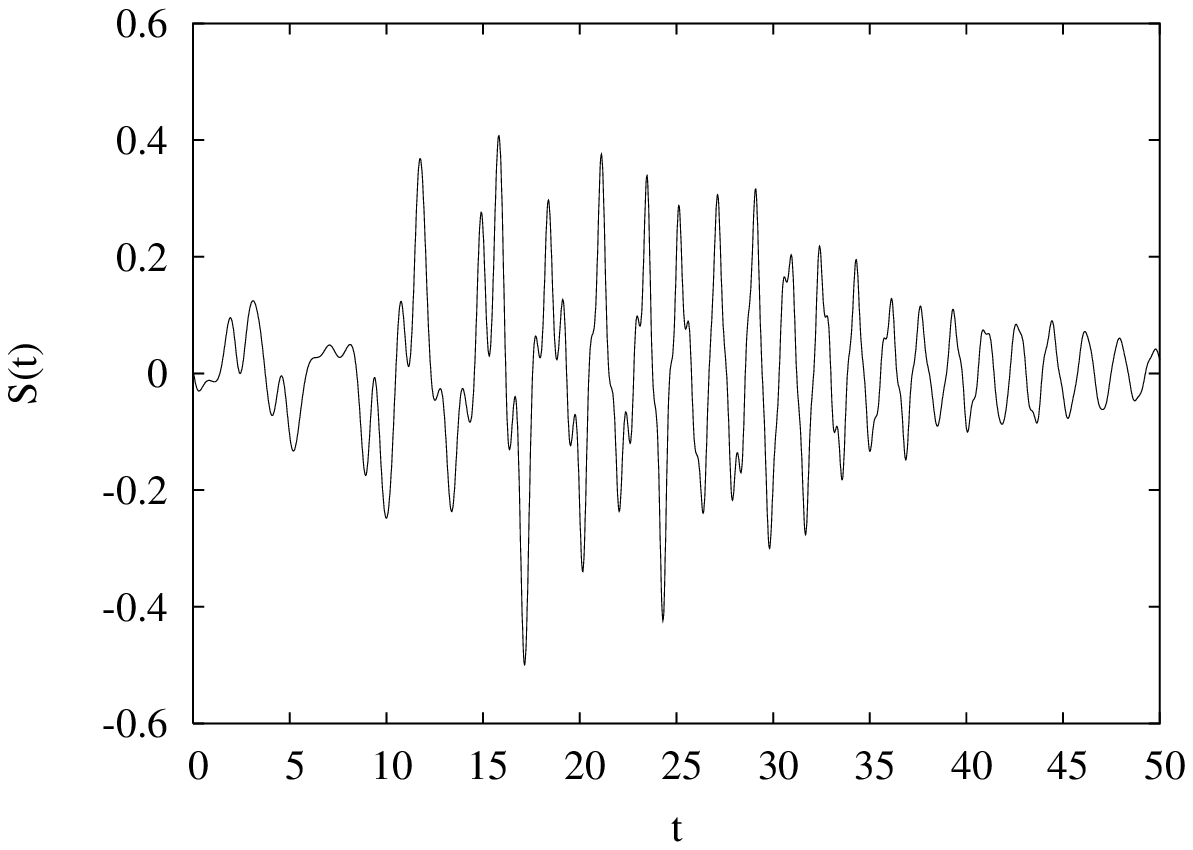}
\hspace{0.6cm}(d)\hspace{-0.5cm}\includegraphics[width=7.5cm,height=4.9cm]{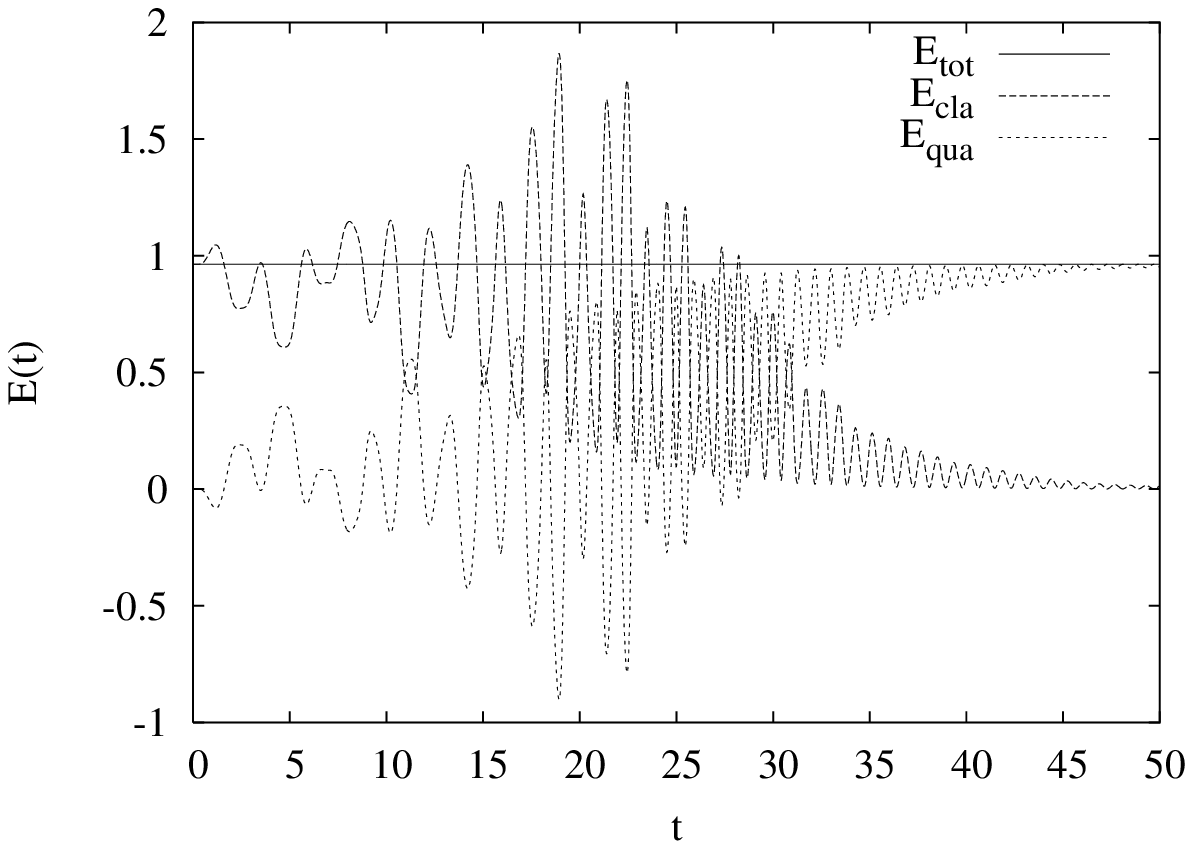}
\end{center}
\caption{\label{fig:sp1.2}Time evolution for the symmetric $\Phi^4$ potential;
Parameters: $m^2=1$ (symmetric potential), $\lambda = 6$,
$\phi(0)=1.2$; (a) evolution of the mean field; (b) evolution
of the effective mass $\calm^2$; (c) evolution of the sunset
contribution to the equation of motion of $\phi$, \eqn{eqphi};
(d) evolution of the energy; in (a-c) the solid lines relate to the
the two-loop 2PPI approximation, the dashed lines to the one-loop or Hartree
approximation; in (d) the dashed line is the classical energy
\eqn{eclass}, the dotted line is the quantum energy.}
\end{figure*}

\begin{figure*}[htbp]
\begin{center}
(a)\hspace{-0.5cm}\includegraphics[width=7.5cm,height=4.9cm]{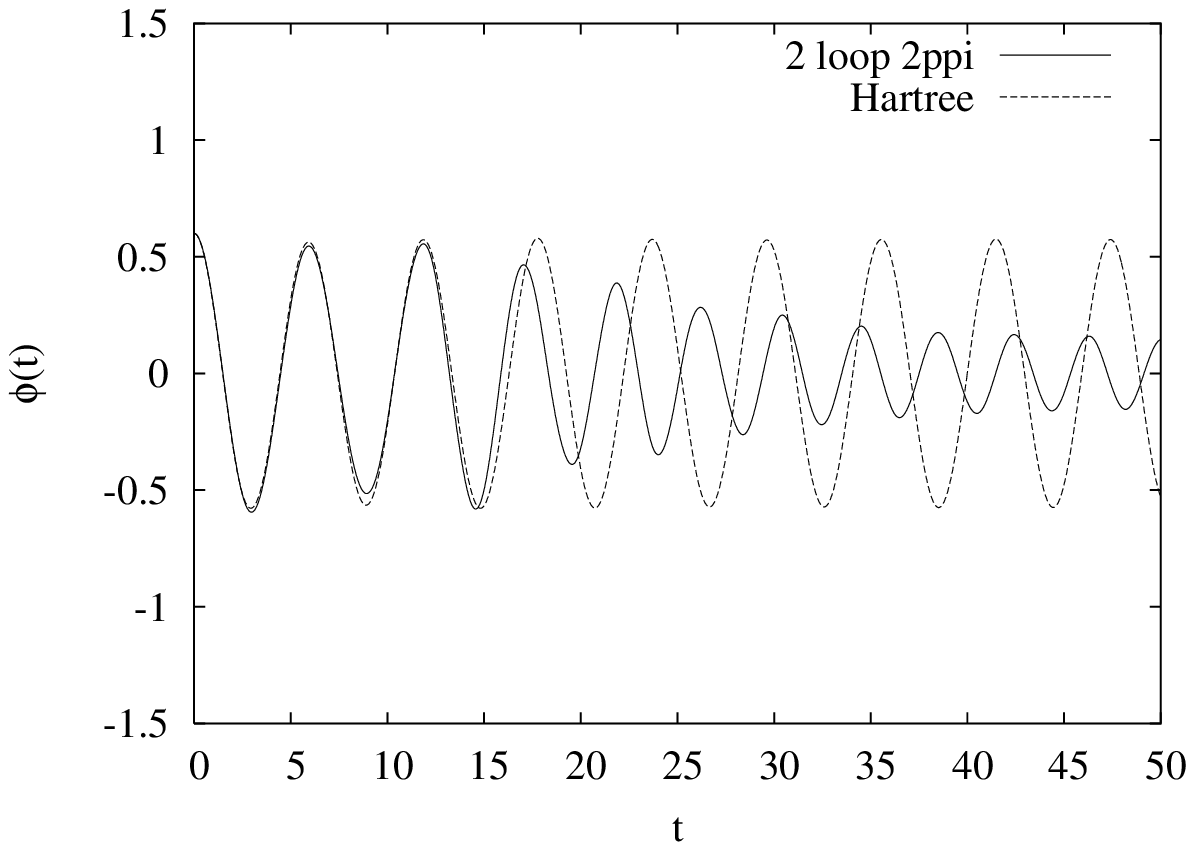}
\hspace{0.6cm}(b)\hspace{-0.5cm}\includegraphics[width=7.5cm,height=4.9cm]{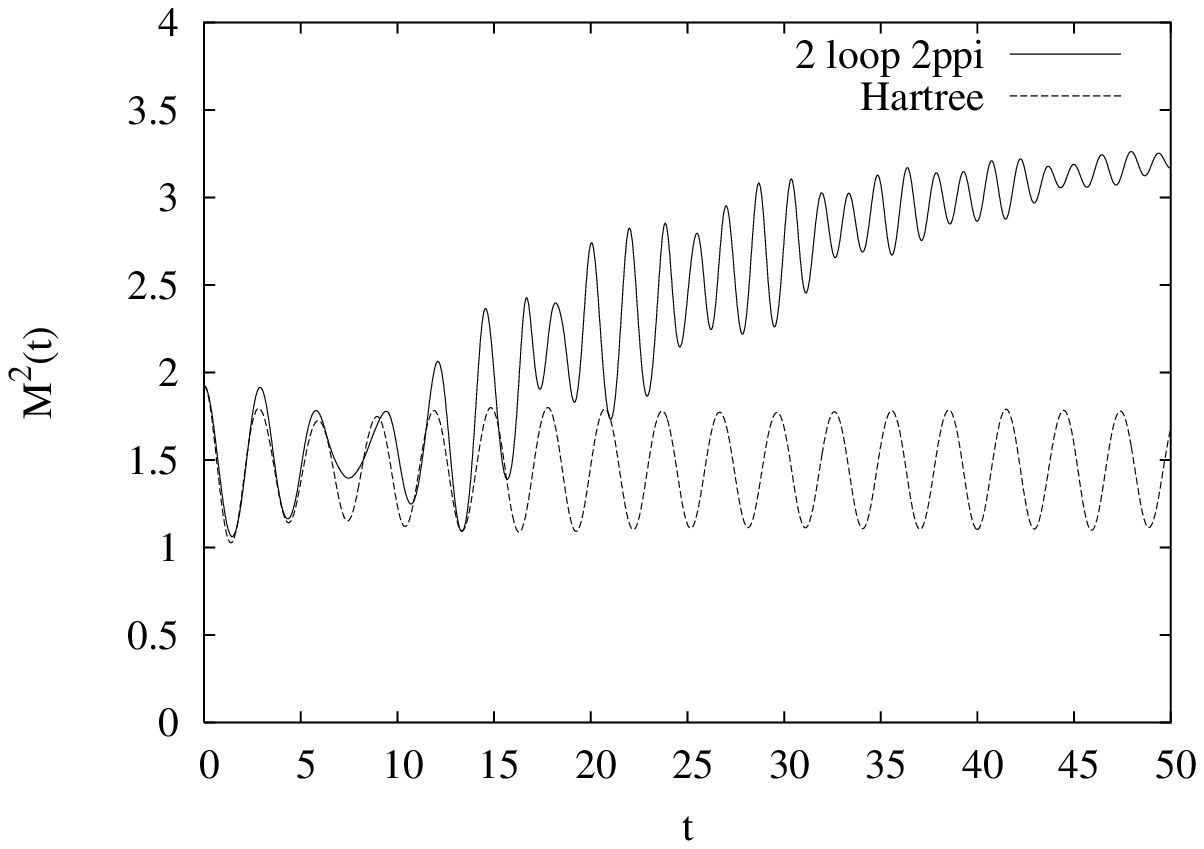}
\vspace{0.2cm}
\par
(c)\hspace{-0.5cm}\includegraphics[width=7.5cm,height=4.9cm]{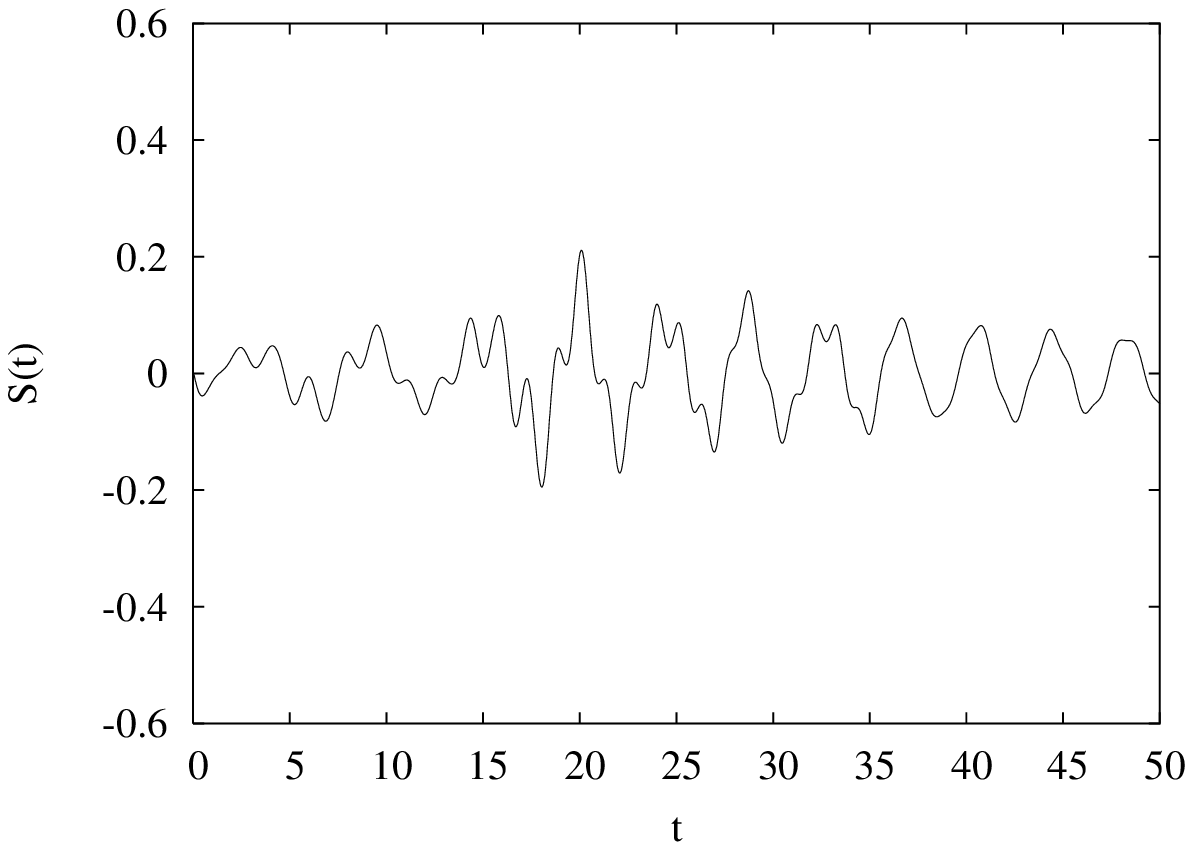}
\hspace{0.6cm}(d)\hspace{-0.5cm}\includegraphics[width=7.5cm,height=4.9cm]{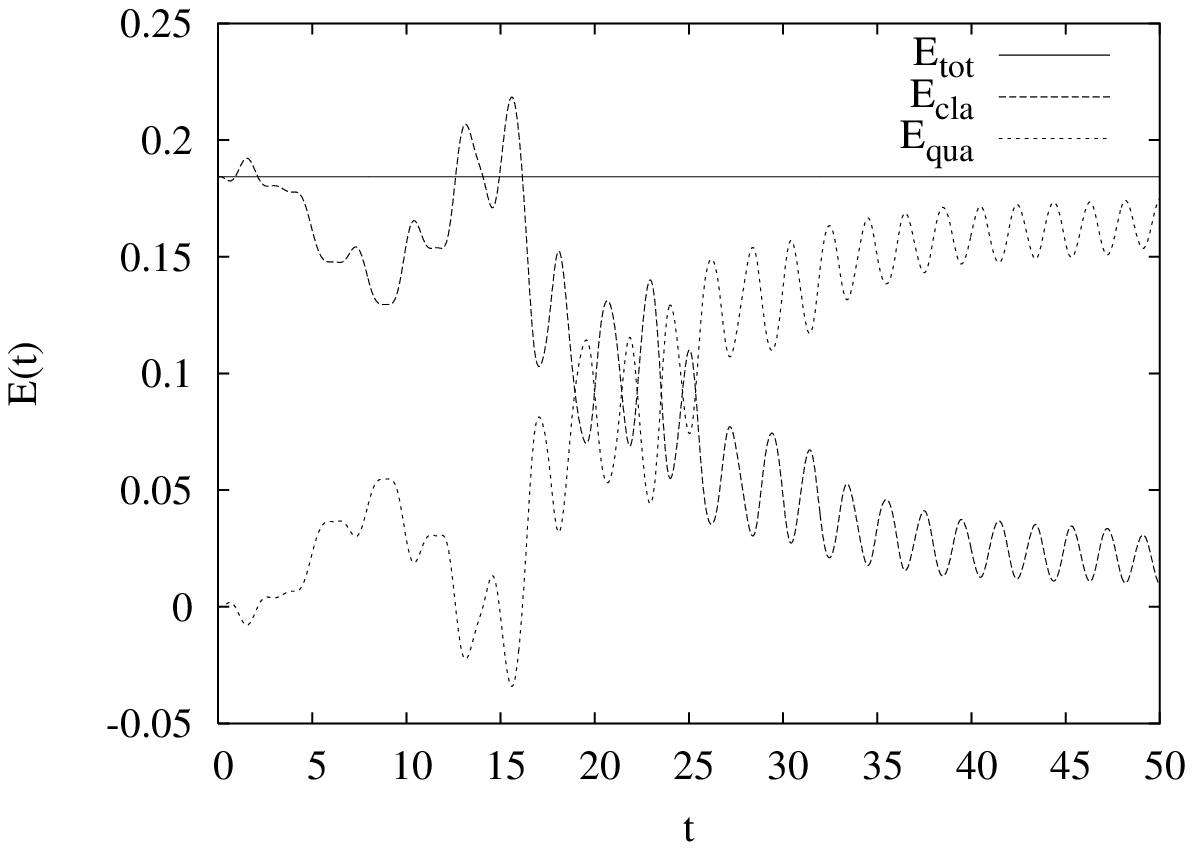}
\end{center}
\caption{\label{fig:sp0.6}
Same as Fig.~\ref{fig:sp1.2} for $\phi(0)=0.6$.}
\end{figure*}

In Figs.~\ref{fig:sp1.2} and \ref{fig:sp0.6} 
we display our numerical results for the time
evolution of the mean field (Figs.~\ref{fig:sp1.2}a and \ref{fig:sp0.6}a), 
of the dynamical mass $\calm^2(t)$ (Figs.~\ref{fig:sp1.2}b and \ref{fig:sp0.6}b),
of the sunset contribution $\cals(t)$ in the classical equation of motion
\eqn{eqphi}  (Figs.~\ref{fig:sp1.2}c and \ref{fig:sp0.6}c) 
and of the classical and quantum parts of the
energy \eqn{etot} (Figs.~\ref{fig:sp1.2}d and \ref{fig:sp0.6}d). 
In the latter diagrams we define the classical 
energy as the standard expression
\be \label{eclass}
E_{\rm cl}=\frac{1}{2}\dot \phi^2 + \frac{1}{2} \left(m^2
+\delta m^2_{\rm fin}\right) \phi^2+
\frac{\lambda}{24}\phi^4
\pkt\ee  
Indeed the repartition between classical and quantum energy
is to some extent arbitrary in a self-consistent framework where,
e.g., $\calm^2$ contains classical as well as quantum parts.
In Figs.~\ref{fig:sp1.2}a,b and \ref{fig:sp0.6}a,b 
we also display the time evolution in the
one-loop or Hartree approximation.

We observe the following characteristic features: after an initial
period of time in which the field amplitude stays roughly constant and
close to the Hartree time evolution a period of effective dissipation sets 
in. For small initial amplitudes evidently the dissipative phase ends
and the mean field reaches a roughly  constant amplitude of oscillation, 
again. For large initial amplitudes such a ``shut off'' is less
evident.  A closer investigation shows that initially the quantum
modes build up until the sunset diagram becomes important. From then on
the Hartree and two-loop evolutions differ substantially. The increase
of the sunset diagram triggers dissipation, until the sunset diagram
again becomes small due to the decrease of the external fields.
Once the sunset diagram has lost its importance the amplitude
of oscillation of the classical field becomes roughly constant again.
This is seen in particular in Fig.~\ref{fig:sp0.6}a, where, due to a relatively
small initial amplitude, the quantum modes and therefore the
sunset diagram are less important than for large initial amplitudes
(or energy densities), as, e.g., in Fig.~\ref{fig:sp1.2}a. We have not followed
the evolution at really large times. So we cannot decide between a constant
and a slowly decreasing amplitude as found, e.g., in the 
large-$N$ case \cite{Boyanovsky:1998zg}. 

The total energy, displayed in Figs.~\ref{fig:sp1.2}d and \ref{fig:sp0.6}d, 
is constant as it should. 
Numerically this is the case within five significant digits or better;
here $E^{(2)}$ was obtained by Runge-Kutta integration of Eq.~(\ref{E2def}).

\begin{figure}[htbp]
\begin{center}
(a)\includegraphics[width=7.5cm,height=4.9cm]{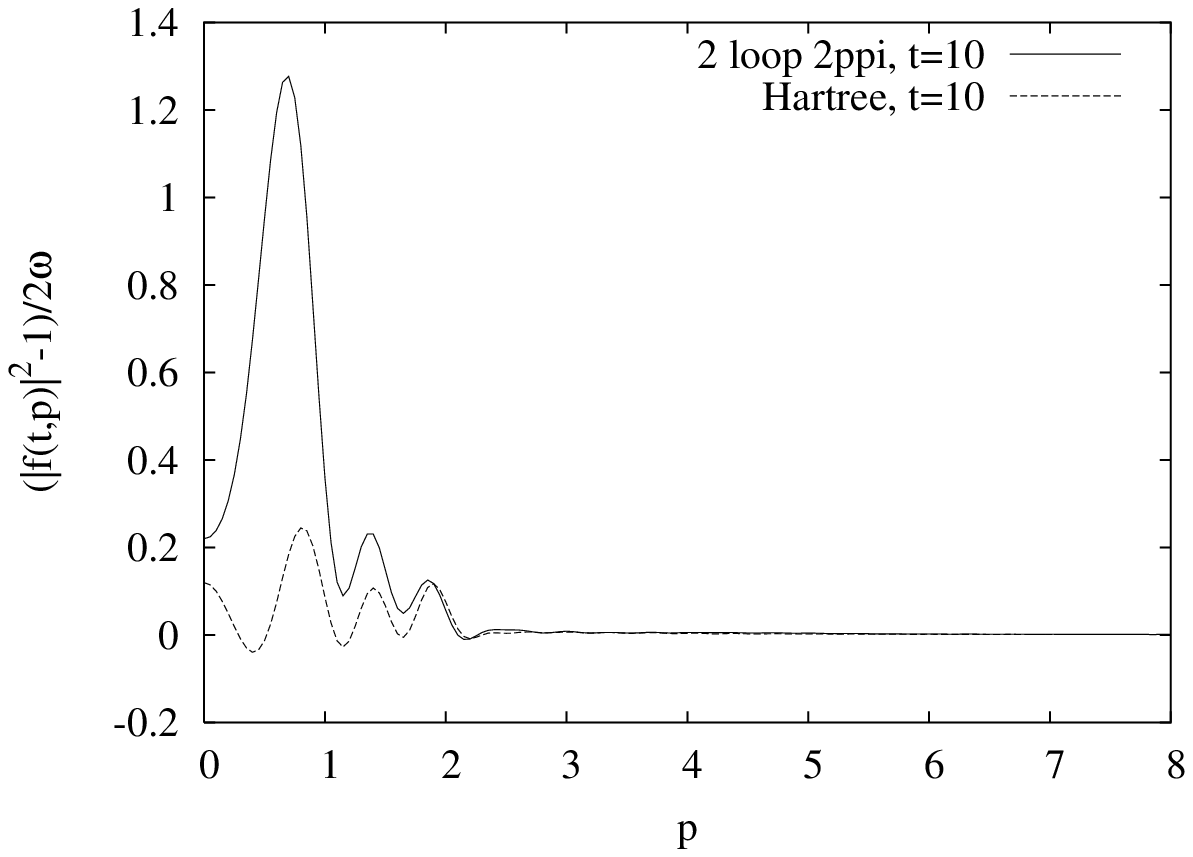}
\par
(b)\includegraphics[width=7.5cm,height=4.9cm]{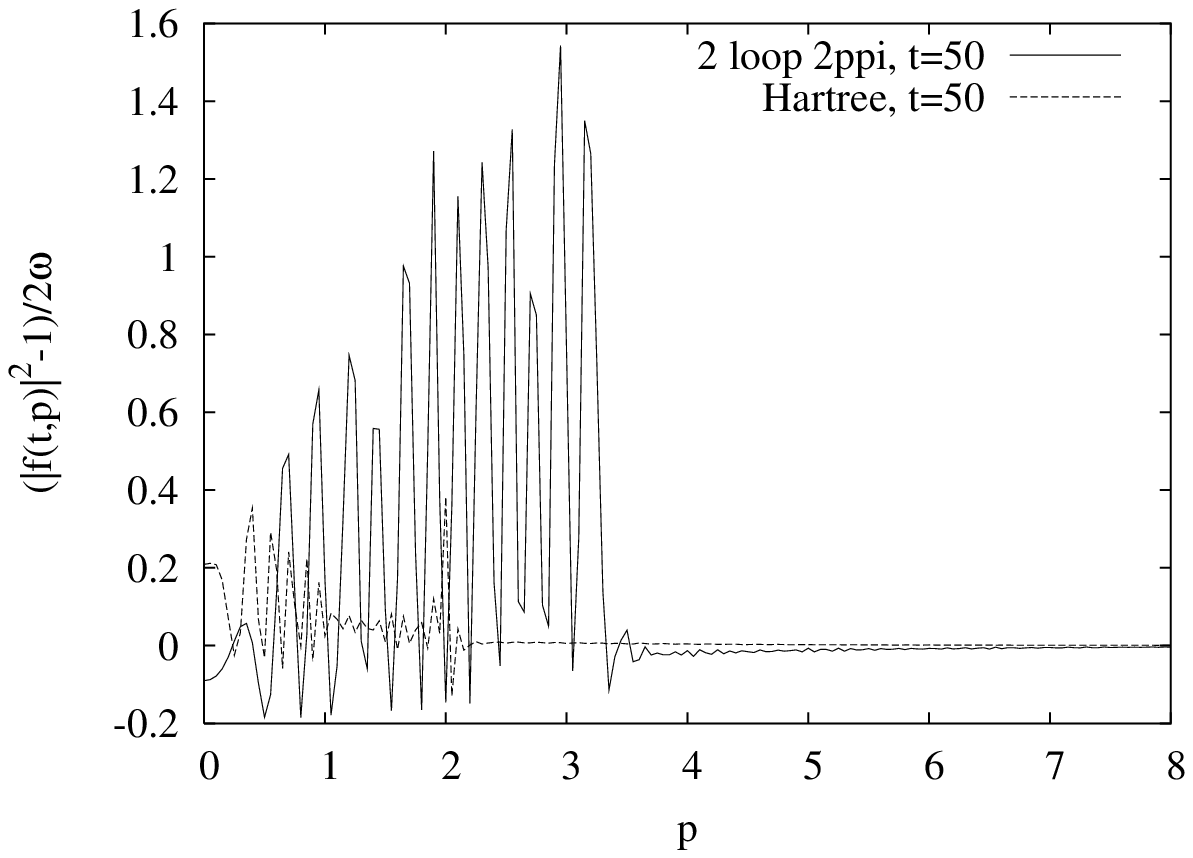}
\end{center}
\caption{\label{fig:spspec}
Momentum spectrum at time $t=10$ and $t=50$
for the parameter set of Fig.~\ref{fig:sp1.2}} 
\end{figure}

We also present, in Fig.~\ref{fig:spspec}, typical momentum spectra.
We have chosen the simulation with $\phi(0)=1.2$ and show the
spectra for an early time $t=10$ and at the end of the simulation.
Along with the results for the two-loop approximation we 
display those for the Hartree approximation. Obviously the spectrum
evolves more strongly for the two-loop approximation. At late times
it shows the typical features of a parametric resonance band
\cite{Traschen:1990sw,Kofman:1994rk,Boyanovsky:1996sq,
Baacke:2001zt}.
Above this band the spectra drop to small values and decrease to zero.
It should be evident that our momentum cutoff of $p_{\rm max}=20$
will be sufficiently high even for the multiple integrals.
On the other hand the numerical integration
cannot take into account the finer details of the spectra, in particular
at later times. Of course to some extent these details are 
washed out if averaged over time. Still, to some extent the finer
details of the time evolution of $\phi(t)$ show some dependence
on the choice of $\Delta p$. However, neither here nor in the case
of the double well potential the qualitative features are affected
by these details. 

We have restricted our presentation to one single coupling parameter
$\lambda = 6$. We have performed simulations for smaller values
of $\lambda$, as well; for such values of $\lambda$ the time evolution is
stretched; the dissipation sets in later and extends
over a larger span of time. For $\lambda=1$ the dispersive phase extends
to typically $t=300$. The general, qualitative,  characteristics of the 
time evolution are similar.

\subsection{Results for the double well potential} 

\begin{figure*}[htbp]
\begin{center}
(a)\hspace{-0.5cm}\includegraphics[width=7.5cm,height=4.8cm]{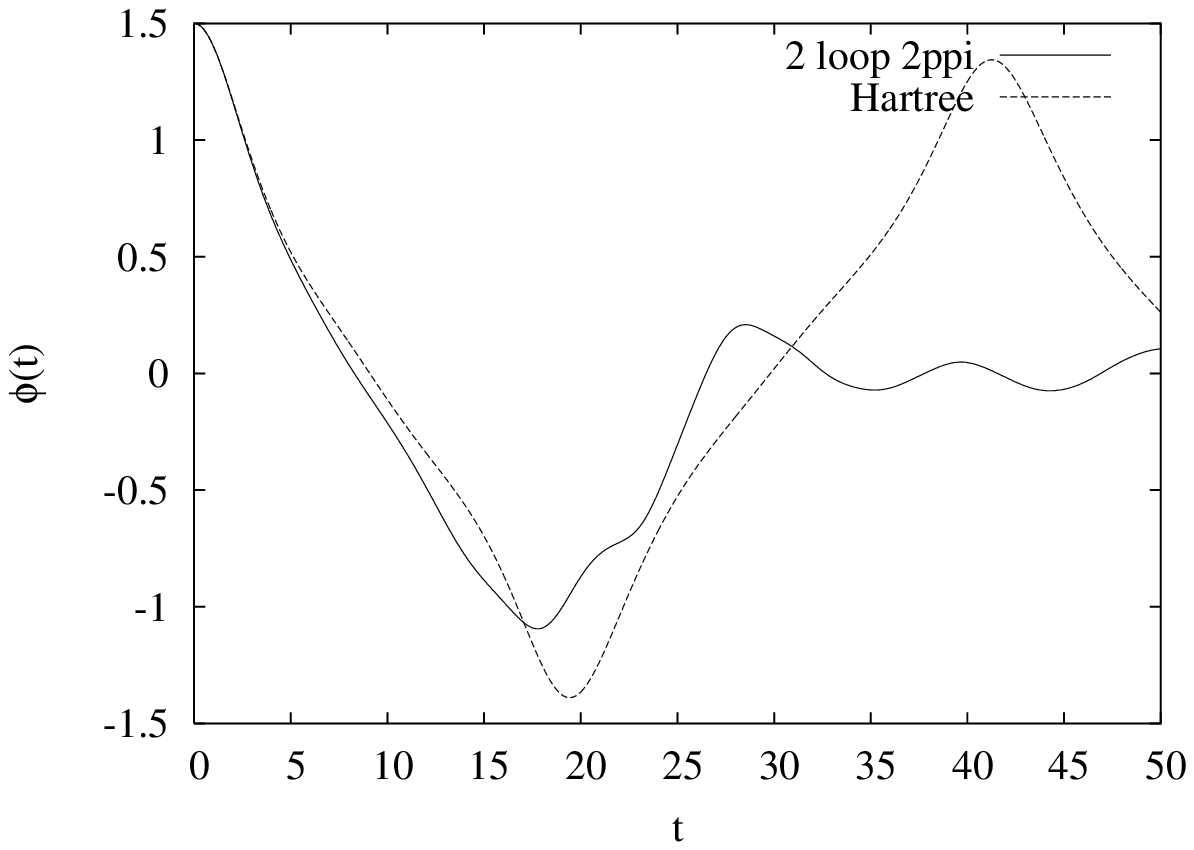}
\hspace{0.6cm}(b)\hspace{-0.5cm}\includegraphics[width=7.5cm,height=4.8cm]{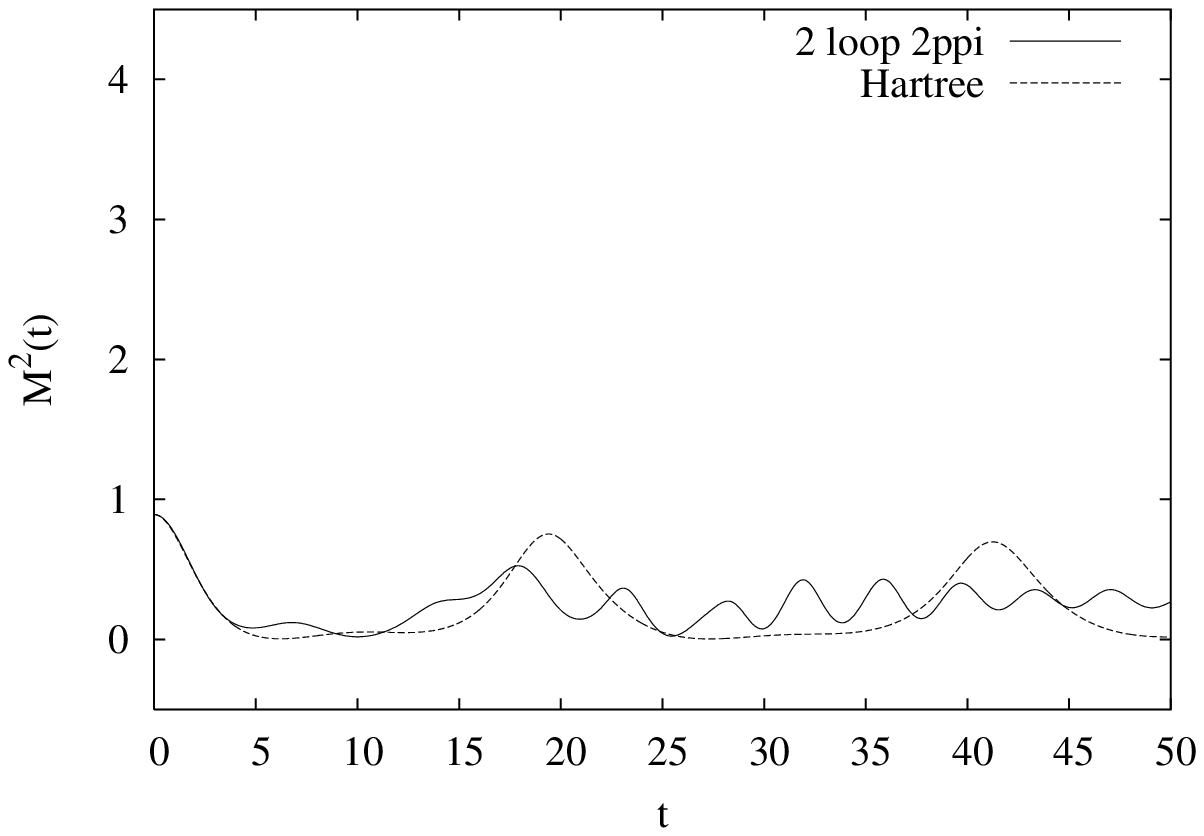}
\vspace{0.2cm}
\par
(c)\hspace{-0.5cm}\includegraphics[width=7.5cm,height=4.8cm]{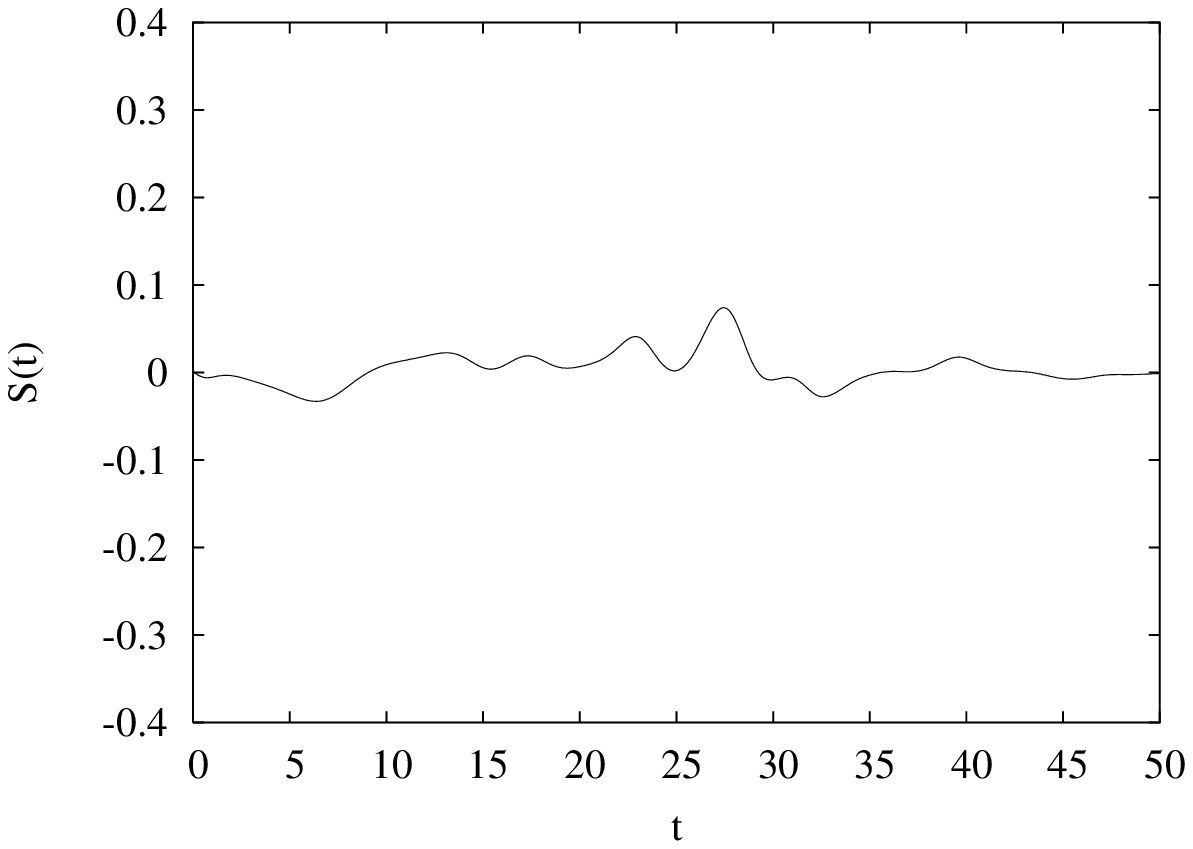}
\hspace{0.6cm}(d)\hspace{-0.5cm}\includegraphics[width=7.5cm,height=4.8cm]{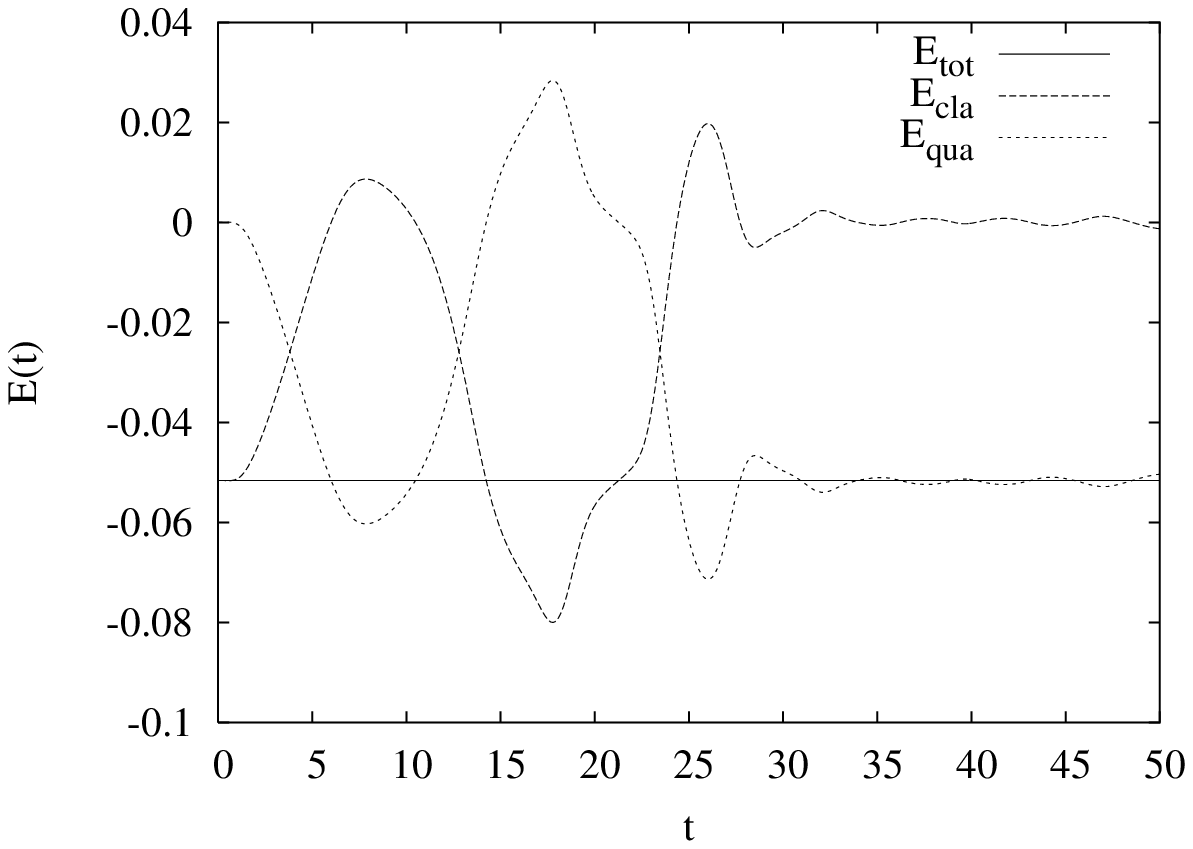}
\end{center}
\caption{\label{fig:dw1.5}
Time evolution for the double well potential.
Parameters: $m^2=-1/6$, $\lambda = 1$, $\phi(0)=1.5$; 
other specifications as in Fig.~\ref{fig:sp1.2}.}
\end{figure*}

\begin{figure*}[htbp]
\begin{center}
(a)\hspace{-0.5cm}\includegraphics[width=7.5cm,height=4.8cm]{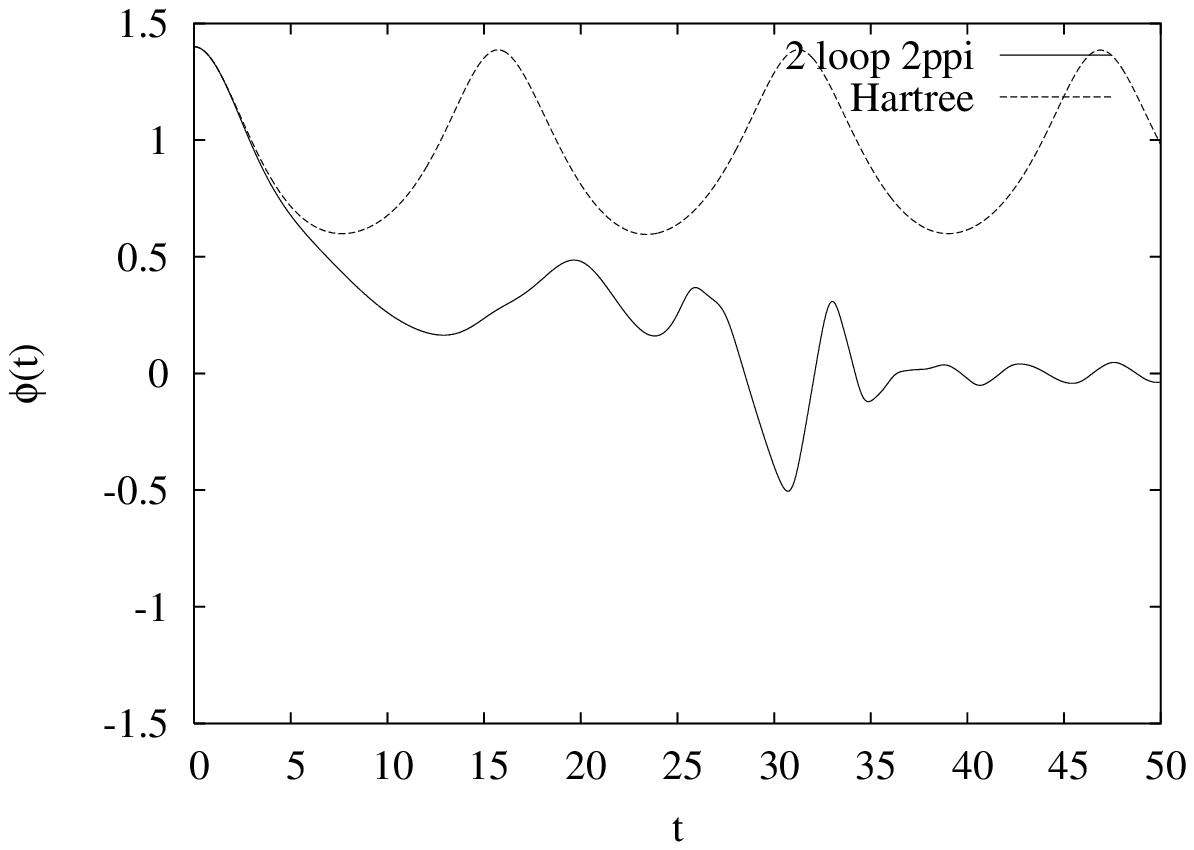}
\hspace{0.6cm}(b)\hspace{-0.5cm}\includegraphics[width=7.5cm,height=4.8cm]{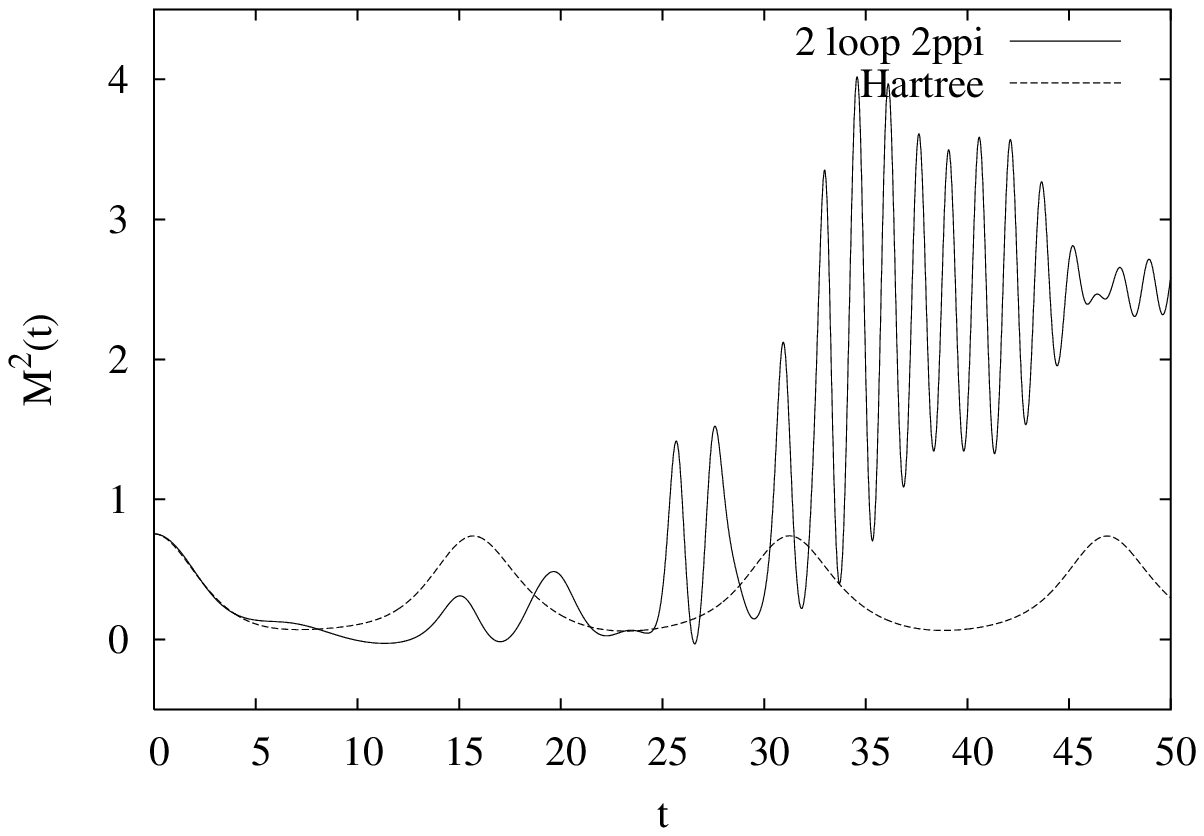}
\vspace{0.2cm}
\par
(c)\hspace{-0.5cm}\includegraphics[width=7.5cm,height=4.8cm]{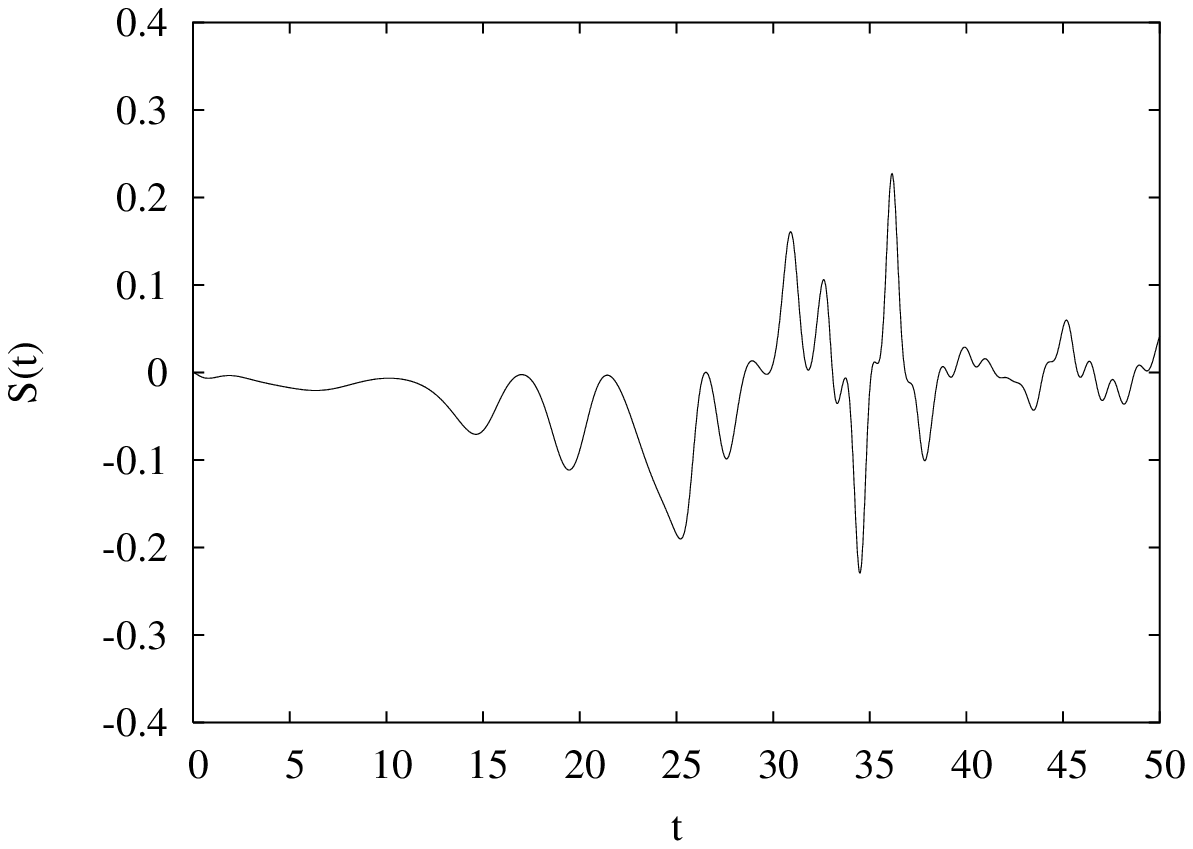}
\hspace{0.6cm}(d)\hspace{-0.5cm}\includegraphics[width=7.5cm,height=4.8cm]{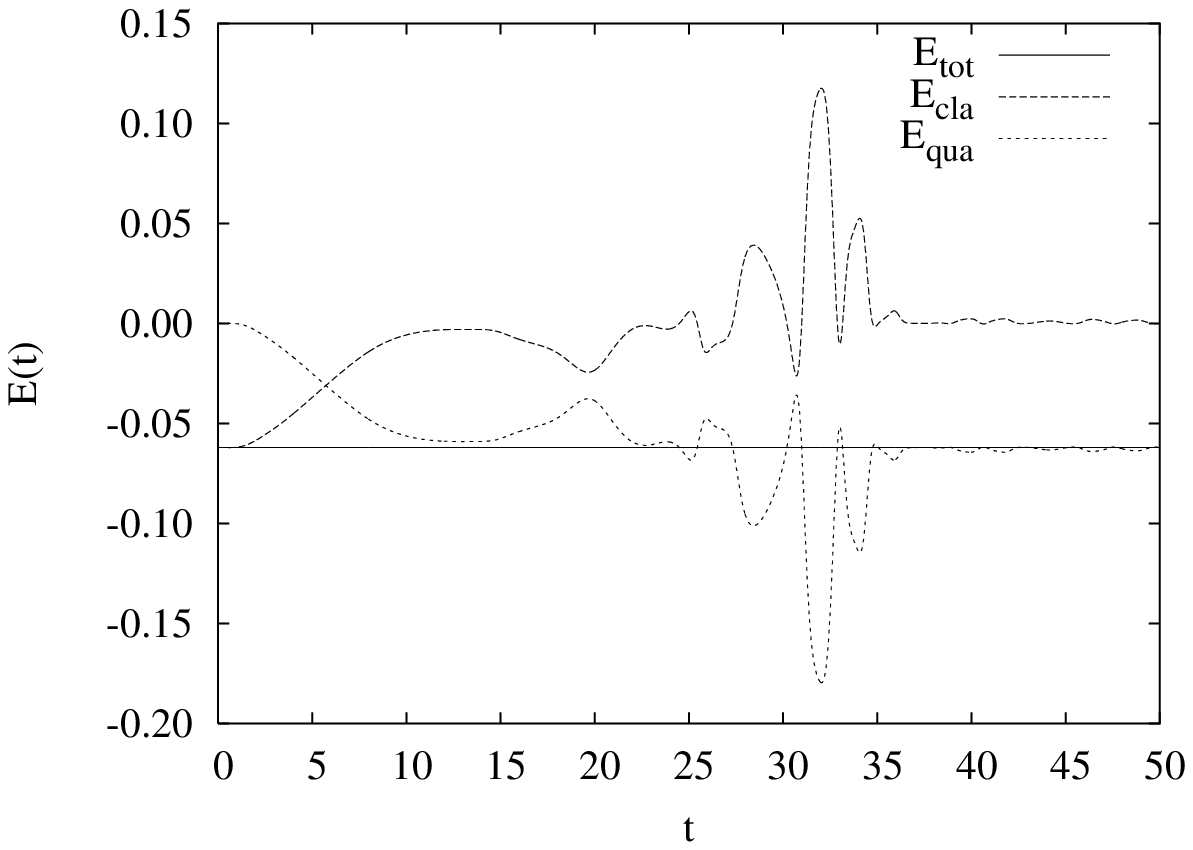}
\end{center}
\caption{\label{fig:dw1.4}
Same as Fig.~\ref{fig:dw1.5} for  $\phi(0)=1.4$.} 
\end{figure*}

\begin{figure*}[htbp]
\begin{center}
(a)\hspace{-0.5cm}\includegraphics[width=7.5cm,height=4.8cm]{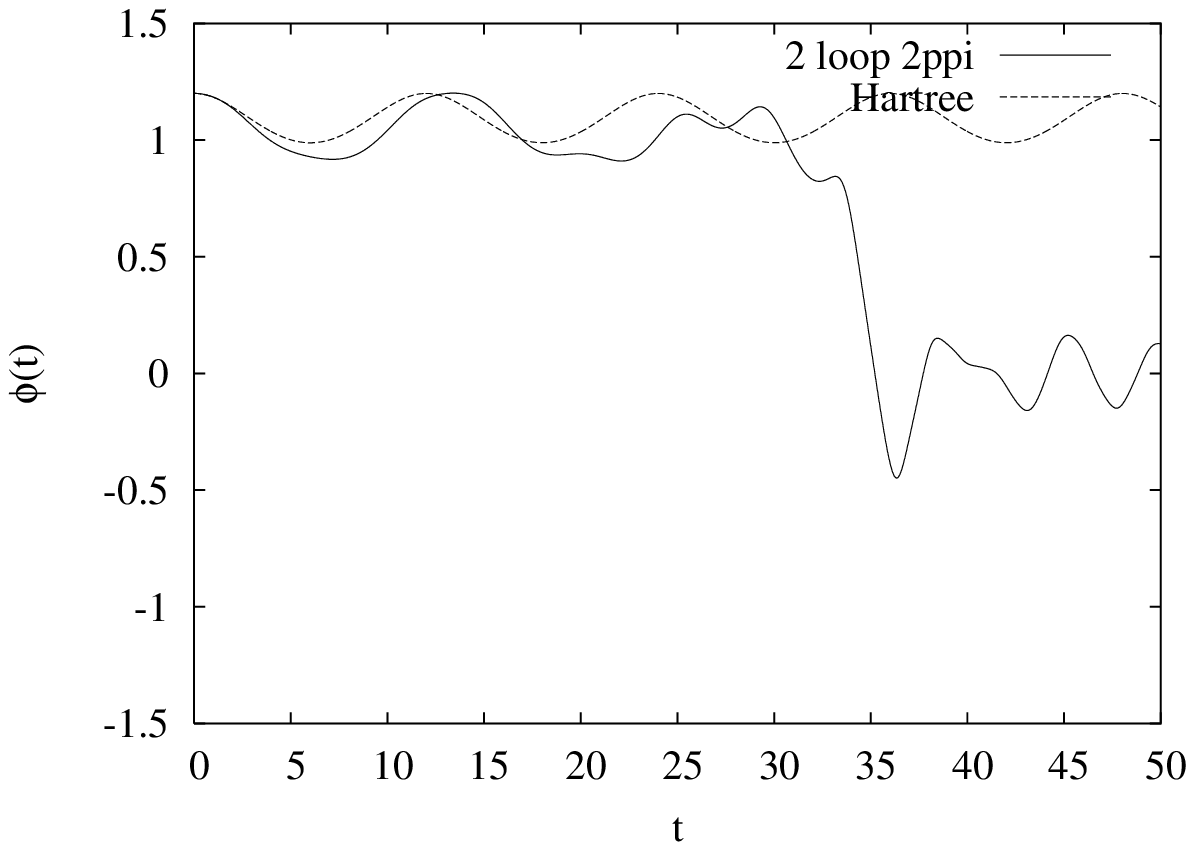}
\hspace{0.6cm}(b)\hspace{-0.5cm}\includegraphics[width=7.5cm,height=4.8cm]{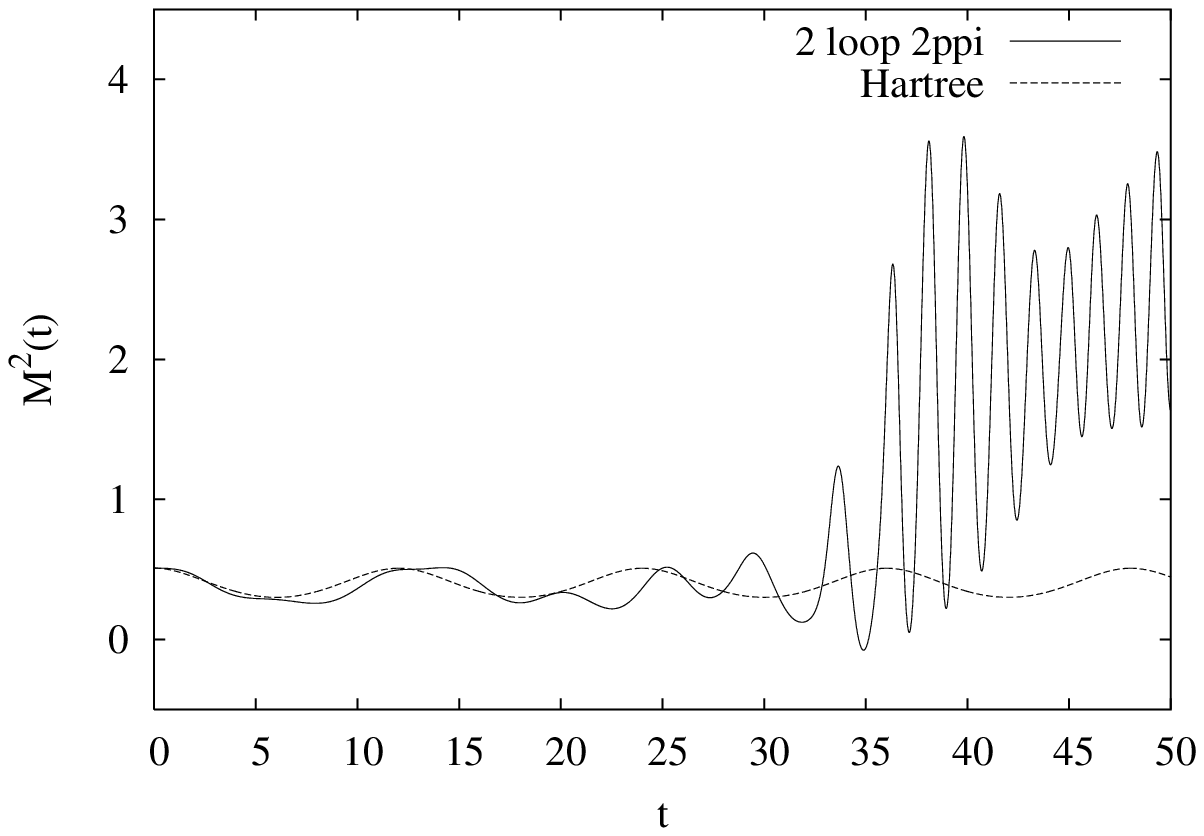}
\vspace{0.2cm}
\par
(c)\hspace{-0.5cm}\includegraphics[width=7.5cm,height=4.8cm]{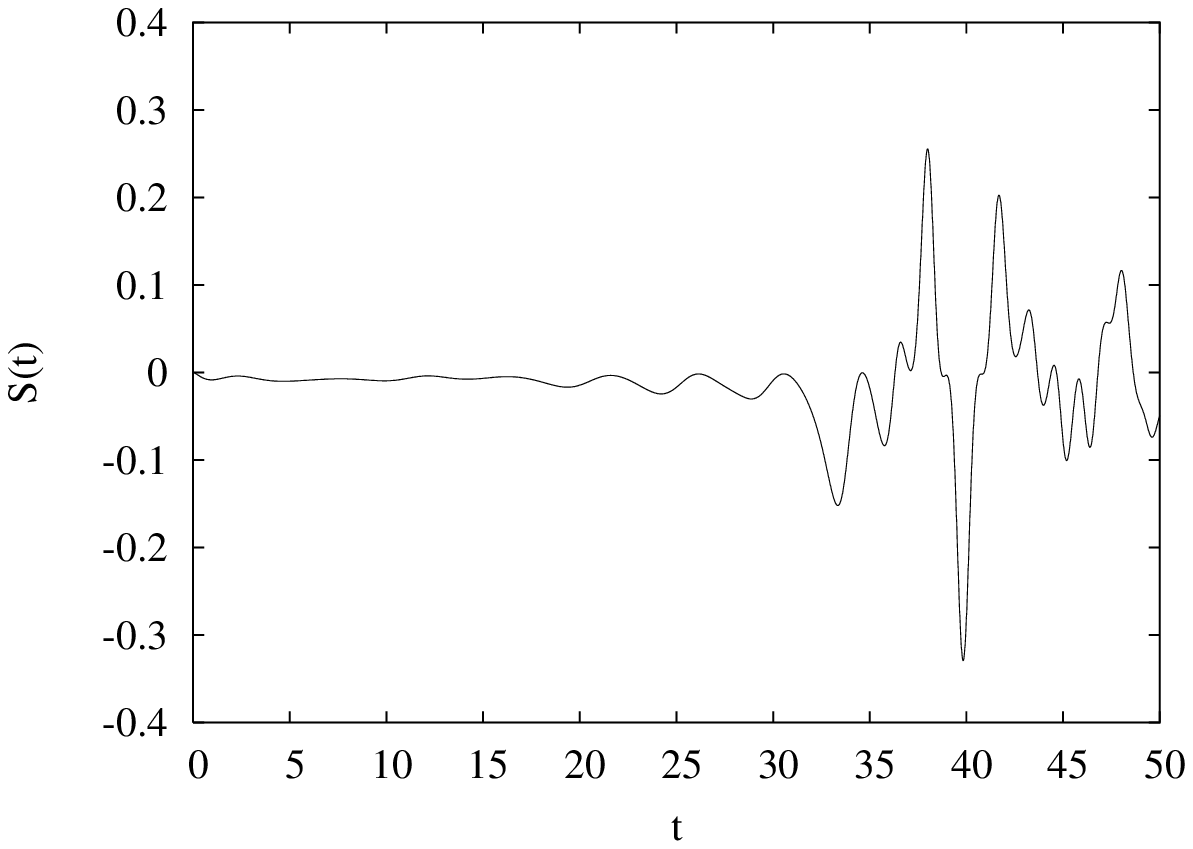}
\hspace{0.6cm}(d)\hspace{-0.5cm}\includegraphics[width=7.5cm,height=4.8cm]{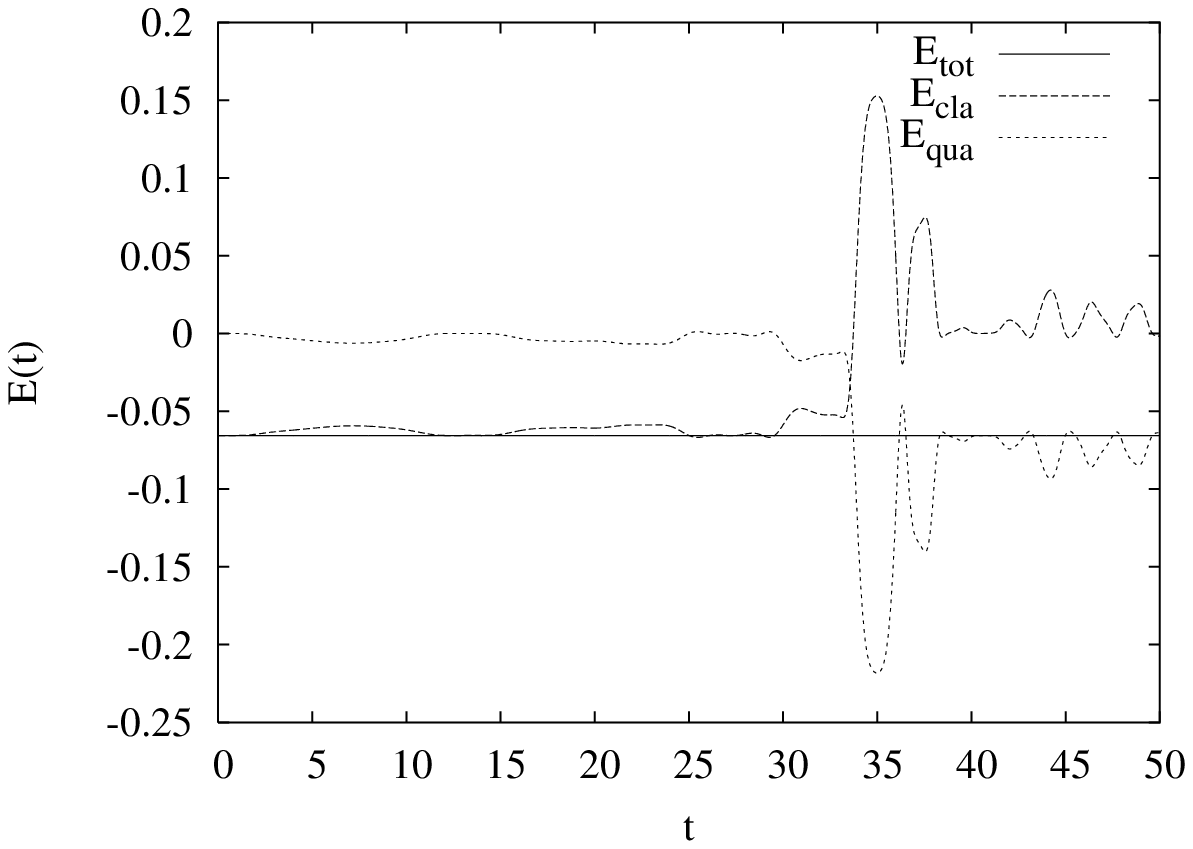}
\end{center}
\caption{\label{fig:dw1.2}
Same as Fig.~\ref{fig:dw1.5} for  $\phi(0)=1.2$.}
\end{figure*}

The numerical simulations for the double well potential
are presented in Figs.~\ref{fig:dw1.5}, \ref{fig:dw1.4} and
\ref{fig:dw1.2}. We take the coupling $\lambda=1$
and $m^2=-1/6$, so that the classical minimum of the double well
potential is at $\phi=v=1$. We consider initial values  $\phi(0) $
equal to $1.5, 1.4$ and $1.2$. Classically the system can
cross the barrier between the two minima for $\phi(0) > \sqrt{2}$.
The first of our initial values is above this critical value, the second
one is slightly below it. In the Hartree approximation
the system evolves as expected from this classical consideration.
In the two-loop 2PPI approximation the system evolves towards
the symmetric phase where the system oscillates around $\phi=0$ 
in the later stages of evolution. The transition between a motion
in the region of the classical minimum and $\phi=0$ is accompanied
by an increase of the sunset contribution and of the mass $\calm^2$.
The transition happens early for $\phi(0)=1.4$. In this case we are
just below the critical value, one sees that the transition towards
$\phi \simeq 0$ happens at a time where the sunset contribution
is still small and where $\calm^2$ only slightly deviates from 
its Hartree value. For $\phi(0)=1.2$ we are deeply in the well.
Here it takes a long time before the evolution towards $\phi 
\simeq 0$ sets in. If we start with values $\phi(0)$ even nearer
to the classical minimum $\phi=1$ the transitions happens 
at even later times and we expect the discretization 
of the momentum spectrum to affect our results so as
to make them unreliable.

\begin{figure}[htbp]
\begin{center}
(a)\includegraphics[width=7.5cm,height=4.8cm]{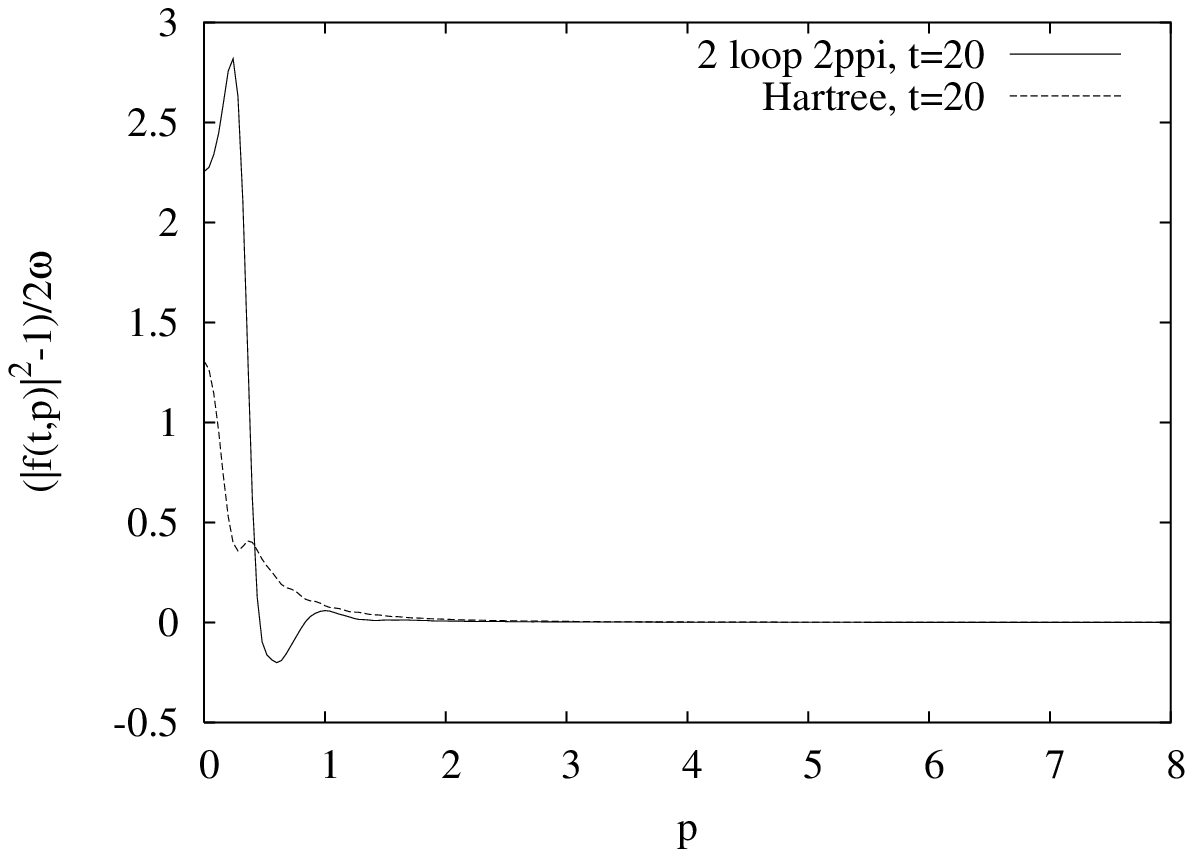}
\par
(b)\includegraphics[width=7.5cm,height=4.8cm]{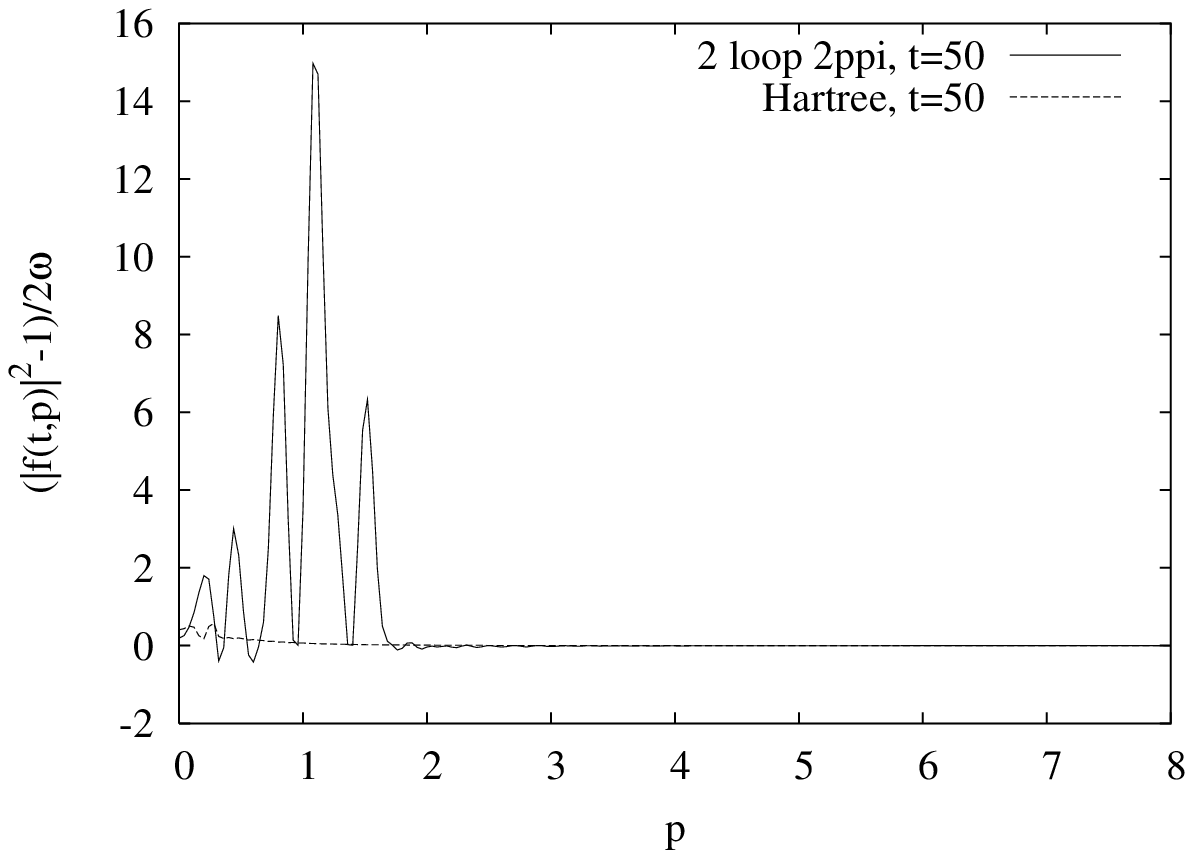}
\end{center}
\caption{\label{fig:dwspec}
Momentum spectrum at time $t=20$ and $t=50$
for the parameter set of Fig.~\ref{fig:dw1.4}}
\end{figure} 

In Fig.~\ref{fig:dwspec} we display momentum spectra
for $|f(t,p)|^2-1$ for the simulation with
$\phi(0)= 1.4$ at $t=20$ and at $t=50$, along with the spectra
obtained in the Hartree approximation. For the two-loop
simulation the spectrum at $t=20$ is characterized by a strong
peak at low momentum, which apparently is due to a passing
of $\calm^2$ to slightly negative values. At $t=50$ the effective
mass of the modes is positive, the spectrum shows a characteristic 
band as typical for parametric resonance.

In all simulations the Hartree approximation displays a rather
clean periodicity which signals a strong coherence between the
evolutions of the classical field and of the quantum modes.
This effect is much stronger than in $3+1$ dimensions, in the Hartree
\cite{Baacke:2001zt} or large-$N$ \cite{Boyanovsky:1996sq,Cooper:1997ii}
approximations.

The nonequilibrium evolution of $\Phi^4$ quantum field theory
with a double well potential has been studied recently by
Cooper et al. \cite{Cooper:2002qd}, with somewhat different
initial conditions. These authors find a transition
towards a symmetric phase in the 2PI formalism extended to
next-to leading order in $1/N$ (2PI-NLO), while in the bare
vertex approximation (BVA) the system remains in the broken phase
for initial energy densities below some critical value.
The exact theory has no phase transition at finite temperature
and, therefore, is not expected to have one at finite energy density
\cite{Griffiths}.

\section{Summary, conclusions and outlook}

In this paper we have  considered the out-of-equilibrium evolution 
of a classical condensate 
field $\phi=\left<\Phi\right>$ and its quantum 
fluctuations for a $\Phi^4$ model
 in $1+1$ dimensions, with a symmetric and a double well potential.
Our investigation was based on the 2PPI formalism in the 
two-loop approximation. We have generalized the 2PPI formalism
to nonequilibrium quantum field theory. In order to find the main
features of this approximation we have performed 
a first set of numerical simulations and compared the 
results to the ones obtained in the Hartree approximation. 
\\\\
We summarize our results as follows:

In the symmetric $\Phi^4$ theory we observe
that the mean field shows a stronger dissipation than the one
found in the Hartree approximation.
The dissipation is roughly exponential in an intermediate time region.
This dissipation is obviously related to the sunset contributions.
As these involve the mean field amplitude they become unimportant
when the amplitude goes to zero. Therefore, for later times the system 
seems to develop a stage of weak dissipation. However,
we have not extended our study to ``late'' times in the sense
of an asymptotic analysis of the evolution. 

In the theory with spontaneous symmetry breaking, i.e., with a double well
potential, the field amplitude tends to zero, i.e., to the symmetric
configuration. This is expected on general grounds: in $1+1$ dimensional 
quantum field theory there is  no spontaneous symmetry breaking
for $T >0$, and so there should be none at finite energy density 
(microcanonical ensemble), either.

We observe in both cases that parametric resonance phenomena are important, 
and that the momentum spectra show no sign of thermalization. In contrast 
to the 2PI approximation the 
interaction between the modes is via the spatially homogenous 
(zero momentum) mass term; so there is no direct momentum exchange
between the modes via a Schwinger-Dyson equation and
the analysis of Ref. \cite{Calzetta:2002ub} concerning thermalization
in the 2PI formalism therefore does not apply.
Our numerical analysis does not allow definite conclusions about
thermalization at later times.

In conclusion we have shown that the 2PPI formalism can be generalized to
nonequilibrium quantum field theory and that the simulations 
in the two-loop approximation in
$1+1$ dimensions show sizeable differences when compared to
the Hartree approximation. Both the stronger dissipation and 
the correct symmetry structure overcome obvious deficits of the
Hartree approximation. We therefore think that it is worthwhile to further
investigate the properties of this approximation, in and out of
equilibrium.

Obvious generalizations of this investigation include the analysis
of an $O(N)$ model with $N>1$ in $1+1$ dimensions and
analogous studies in $3+1$ dimensions. The technical requirements
for such simulations are considerably reduced when compared to 
the 2PI formalism in the analogous approximation, due to the
factorization of the Green functions; moreover the problem of 
renormalization in $3+1$ dimensions has been solved in equilibrium
\cite{Verschelde:2000ta,Verschelde:2000dz}. In the 2PI approach
the three-loop renormalization has been considered in
\cite{vanHees:2001ik} for the mean field $\phi=0$ case;
an analysis of renormalization beyond the Hartree
approximation is still lacking for $\phi\neq 0$.

We feel that it is very important to accompany the 
numerical simulations of nonequilibrium systems in various
formalisms and approximations by equivalent analyses
for systems in thermal equilibrium. Such analyses are still
lacking entirely.


\section*{Acknowledgments}
The authors take pleasure in thanking Stefan Michalski, Hendrik van Hees
and Henri Verschelde for useful and stimulating discussions and the
Deutsche Forschungsgemeinschaft for financial support under contract Ba703/6-1.

\renewcommand{\theequation}{\Alph{section}.\arabic{equation}}


\appendix 

\setcounter{equation}{0}

\section{Comparison between 2PI and 2PPI}
\setcounter{equation}{0}
In this section we are giving some comments on the differences between
the 2PI (two particle irreducible) and the 2PPI (two particle
\emph{point} irreducible) formalism at the two-loop level.

The 2PI effective action reads \cite{Cornwall:1974vz}
\bea
\Gamma[\phi,G]&=&S[\phi]+\frac{1}{2}i\mathrm{Tr}\ln
G^{-1}+\frac{1}{2}i\mathrm{Tr}\left(D^{-1}G\right)\nonumber \\
&&+\Gamma_2[\phi,G] \ ,
\eea 
where
$iD^{-1}(x,x')=-(\square+m^2)\delta(x-x')-\frac{\lambda}{2}\phi^2(x)\delta(x-x')$
is the classical propagator.
If we truncate $\Gamma_2$ including two-loop order terms we have
\bea
\Gamma_2^{(2)}[\phi,G]=\includegraphics[scale=0.25]{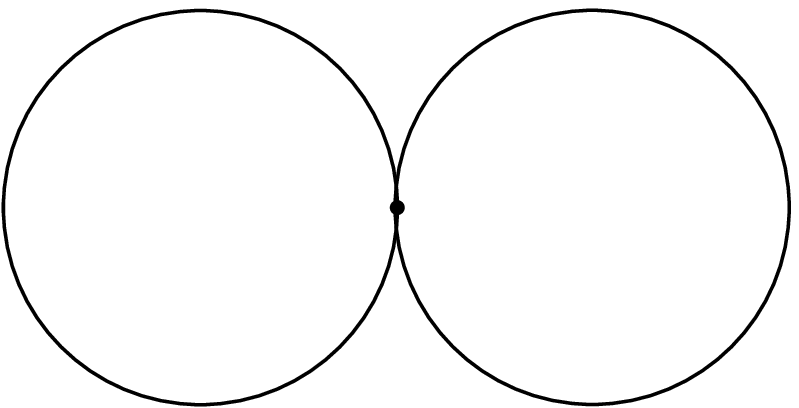}\
+\ \includegraphics[scale=0.25]{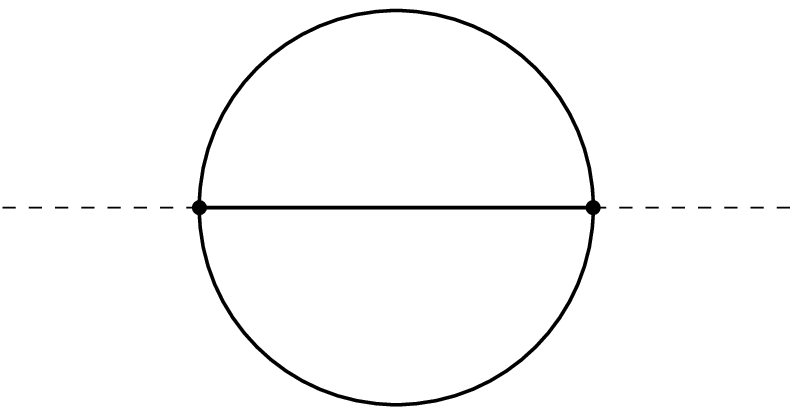} \ .
\eea
A variation of $\Gamma_2^{(2)}$ with respect to $G$ gives the self
energy
\bea
\Sigma^{(2)}(x,x')&=&2i\frac{\delta \Gamma_2^{(2)}[\phi,G]}{\delta
  G(x,x')}\nonumber \\
&=&\includegraphics[scale=0.25]{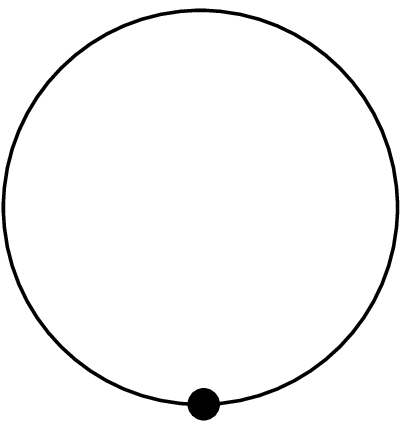}\
+\ \includegraphics[scale=0.25]{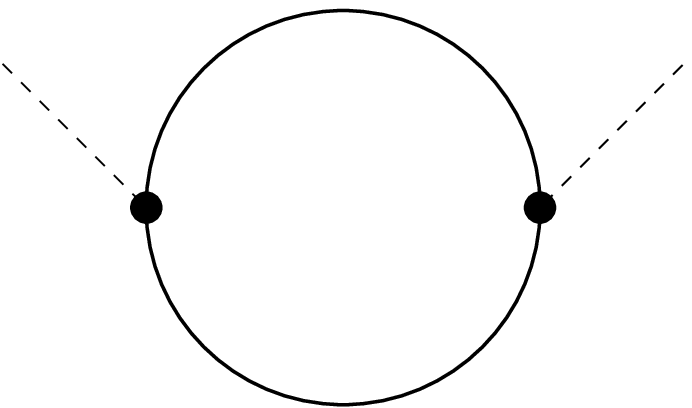} \ .
\eea
The two point function $G$ fulfills the Schwinger-Dyson equation 
\cite{Cornwall:1974vz}
\bea
iG^{-1}(x,x')=iD^{-1}(x,x')-i\Sigma^{(2)}(x,x') \label{eq:2pi-SD}\ 
\eea
and is a variational parameter of the formalism.

The relevant formula for the 2PPI formalism in the two-loop
approximation are given in section~\ref{subsec:sunset}. 
Note that although the sunset contribution to the effective
action in both the 2PI and 2PPI approach is depicted by the same
diagram, they have different implications. We think that it is instructive to
compare some implicit 1PI graphs of both approximations to emphasize
the differences. These 1PI graphs are hidden in the resummation and arise 
via the self consistent Schwinger-Dyson or gap equation of the 2PI
(see Eq.~(\ref{eq:2pi-SD}))
or 2PPI
formalism (see Eq.~(\ref{eq:M2cal})), respectively.

\begin{figure}[htbp]
  \centering
  \includegraphics[scale=0.5]{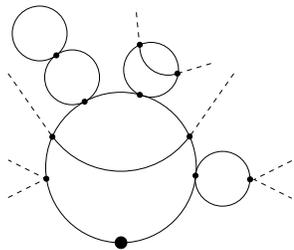}
  \caption{  \label{fig:gen-2ppi}
 A generic 1PI but 2PPR graph, which is produced via the resummation
  in the gap equation for $\calm^2(t)$ in the two-loop approximation
  in the 2PPI formalism. Thin solid lines denote the
  free propagator while dashed lines denote the classical field
  $\phi$. The solid dot indicates the external time $t$.}
\end{figure}

\begin{figure}[htbp]
  \centering
  \includegraphics[scale=0.5]{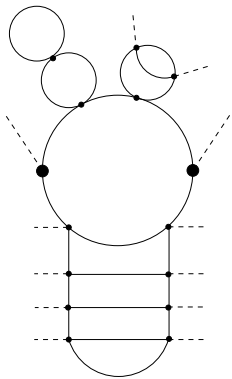}
  \caption{\label{fig:gen-2pi}
A generic 1PI but 2PR graph, which is hidden in the
  resummation via the full Schwinger-Dyson equation~(\ref{eq:2pi-SD}) for the self
  energy $\Sigma^{(2)}(t,t';p)$. Thin solid lines denote the
  free propagator while dashed lines denote the classical field
  $\phi$. The solid dots indicate the external times $t$ and $t'$.}
\end{figure}

\begin{figure}[htbp]
  \centering
  \includegraphics[scale=0.5]{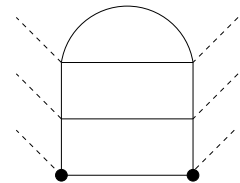}
  \caption{\label{fig:ladder-2pi} Example for a ladder diagram. 
Thin solid lines denote the
  free propagator while dashed lines denote the classical field
  $\phi$. The solid dots indicate the external times $t$ and $t'$.}
\end{figure}

In Fig.~\ref{fig:gen-2ppi} we present such a generic 1PI but 2PPR graph in the
two-loop 2PPI approximation. In Fig.~\ref{fig:gen-2pi} we display a
similar graph in the two-loop 2PI approach. As the 2PPI formalism
resums all \emph{local} contributions to the propagator \emph{no ladder diagrams}
are introduced via resummation. In the 2PI formalism in addition nonlocal insertions
are taken into account which lead to infinite ladder resummations. An
example for such a ladder diagram is depicted in
Fig.~\ref{fig:ladder-2pi}. It can be identified in the lower part of
Fig.~\ref{fig:gen-2pi}. 

As ladder diagrams do not fall apart if two lines meeting at the same
point are cut, they are indeed 2PPI and thus join in the 2PPI formalism
explicitly as higher order corrections to the effective action
functional $\Gamma$. We have shown a three-loop diagram of
``ladder-type'' in Fig.~\ref{fig:2PPR2PR}b. 

\begin{figure}[htbp]
  \centering
(a)  \includegraphics[scale=0.5]{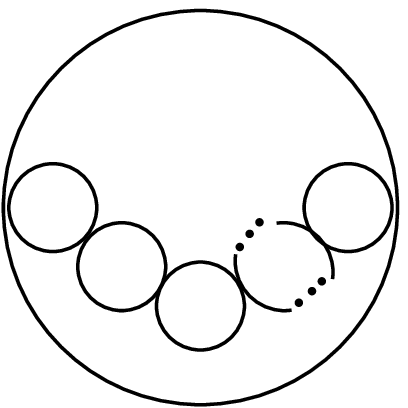} \qquad
(b)  \includegraphics[scale=0.5]{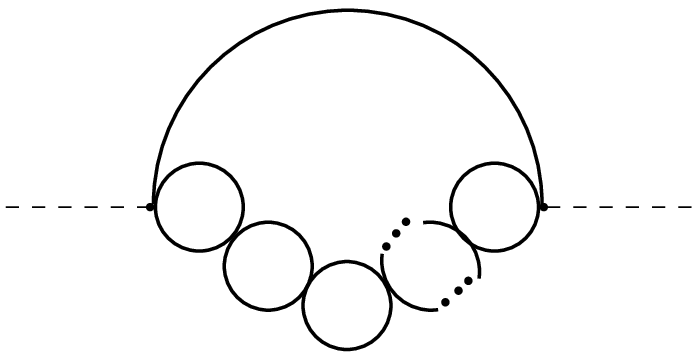}
  \caption{\label{fig:chains-2pi} Vacuum graphs with a topology of
    closed chains which contribute to
    the 2PI effective action in the NLO-1/N approximation; solid lines
    represent the 2PI propagator $G$, while the dashed lines denotes
    the classical field $\phi$. }
\end{figure}

The BVA and NLO-1/N approximations in the 2PI formalism sum an even
larger class of diagrams as in these approximations already 
$\Gamma^\mathrm{2PI}[\phi,G]$ contains an
infinite series of vacuum diagrams with all loop orders. This infinite series of
chain diagrams can be formulated in a
very compact way within the auxiliary field formalism
\cite{Blagoev:2001ze, Aarts:2002dj}. The diagrams have the 
topology of chains of bubble-graphs (see
Fig.~\ref{fig:chains-2pi} for two generic vacuum graphs contributing
to $\Gamma^\mathrm{2PI}$). Depending on a given approximation these
diagrams contribute in the 2PPI formalism as well. 

\section{Some more comments on the numerics and momentum integrations}
\setcounter{equation}{0}
In our simulations we have used a momentum cutoff of
$p_{\rm max}=20$ and an equidistant momenum grid with
$\Delta p = 0.05$. Both choices are somewhat generous; we have
not attempted an optimization with respect to CPU time and
storage requirements as one would certainly have to do for
simulations in $3+1$ dimensions. In this Appendix we more closely
investigate the cutoff and momentum grid dependences. 

\begin{itemize}
\item[(i)] The choice of $\Delta p$.
We have repeated the simulation of
Fig.~\ref{fig:dw1.4} for values of $\Delta p$ between $0.04$ and $0.2$ while leaving
$p_{\rm max}=20$ fixed. We display in Fig.~\ref{fig:mom-grid} the time evolution of the
classical field $\phi(t)$ and of the effective mass $\calm^2(t)$. The numerical
results for these quantities are seen to converge for $\Delta p \lesssim 0.07$. The curves for $\Delta p =0.067$ and  $\Delta p =0.04$
cannot be distinguished. While the qualitative behavior of
$\phi(t)$ does not change even for larger values of $\Delta p$ the 
late time averages of the mass $\calm^2$ show a considerable dependence
beyond $\Delta p \simeq .07$.

\begin{figure*}[htbp]
  \centering
(a)\hspace{-0.5cm}\includegraphics[width=7.5cm]{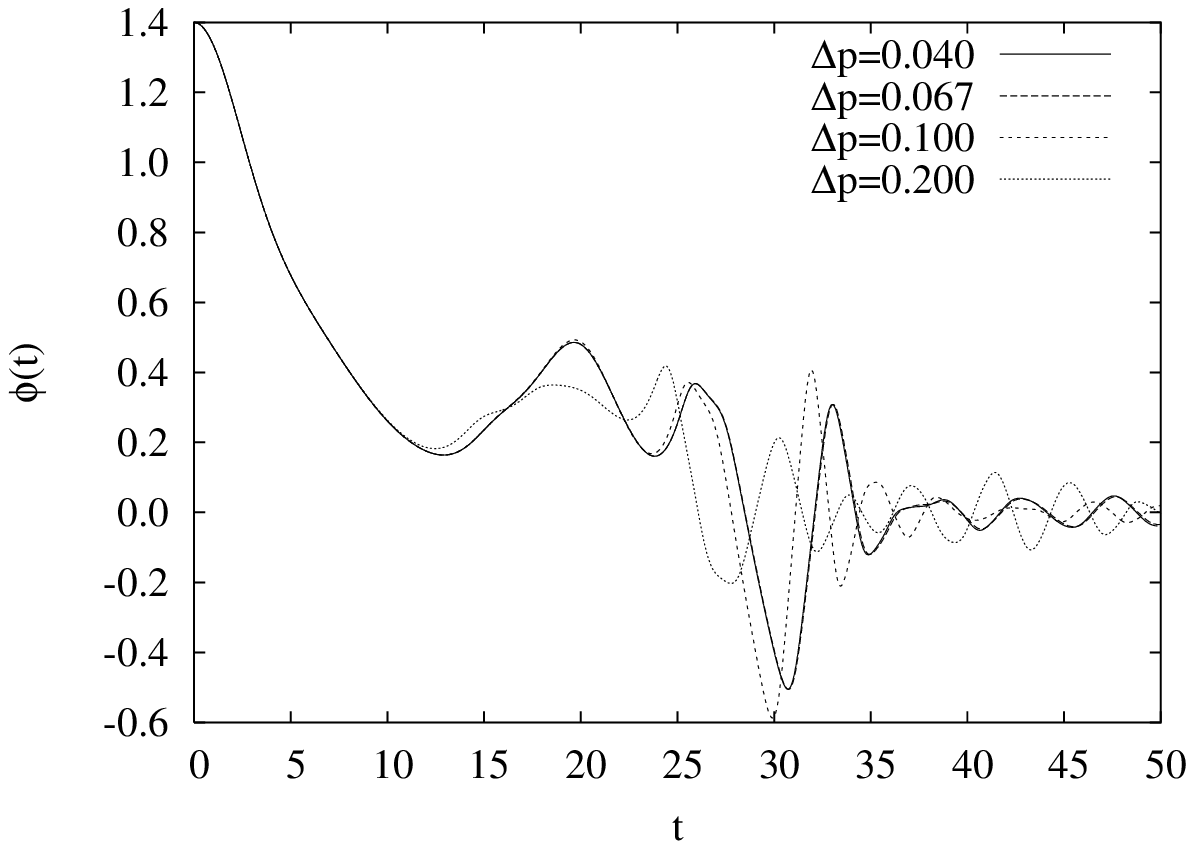} \hspace{0.6cm}
(b)\hspace{-0.5cm}
\includegraphics[width=7.5cm]{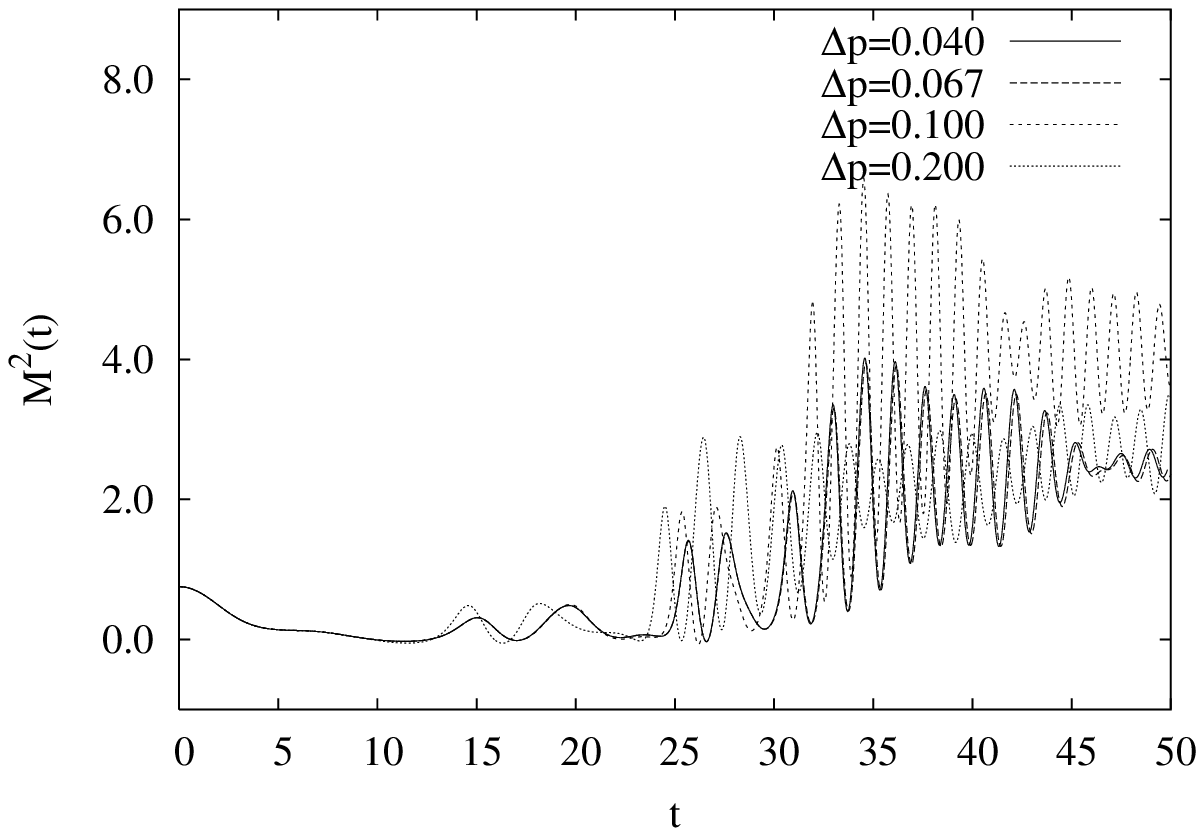}
\caption{\label{fig:mom-grid}Detailed study of the dependence on the
  momentum grid for the simulation with the parameters from
  Fig.~\ref{fig:dw1.4} (a) evolution of the
  mean field $\phi$ (b) evolution for the effective mass $\calm^2$;
  the momentum cutoff is fixed at $p_\mathrm{max}=20$ and we vary
  the distance between the grid points $\Delta p$; the solid line
  represents $\Delta p=0.04$, the long dashed line $\Delta p=0.067$,
  the short dashed line $\Delta p=0.1$ and the dotted line $\Delta p=0.2$.}
\end{figure*}

\item[(ii)] The cutoff dependence. The momentum cutoff is a cutoff 
for convergent integrals. As one may conclude already from
the momentum spectra displayed in Fig.~\ref{fig:dwspec} the cutoff can 
be reduced appreciably. In Fig.~\ref{fig:mom-max} we show the dependence of
$\calm^2(t)$ on $p_{\rm max}$. One sees that even for a cutoff as low
as $p_{\rm max}=5$ the deviations are only at the percent level and
for  $p_{\rm max}=5$ the results are already satisfactory. This may change
at later times if the momentum distributions get broader by rescattering.

\begin{figure}[htbp]
  \centering
\hspace{-0.5cm}\includegraphics[width=7.5cm]{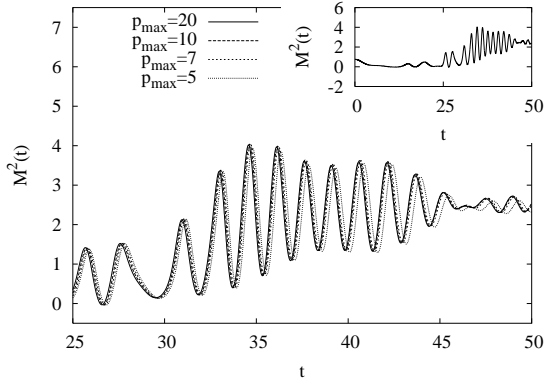}
\caption{\label{fig:mom-max}Dependence of the time evolution of
  $\calm^2(t)$ on the
  momentum cutoff $p_\mathrm{max}$ for the simulation with the parameters from
  Fig.~\ref{fig:dw1.4} in the time range
  $t\in[25,50]$; in the inset the whole time range is shown; the momentum distance
  is fixed at $\Delta p=0.05$ while $p_\mathrm{max}$ varies between
  $5$ and $20$; the solid line represents the simulation for
  $p_\mathrm{max}=20$, the long dashed line $p_\mathrm{max}=10$, the short
  dashed line $p_\mathrm{max}=7$ and the dotted line $p_\mathrm{max}=5$.}
\end{figure}

\item[(iii)]The time grid. For our simulations we have chosen
$\Delta t=0.001$, except for the simulation in Fig.~\ref{fig:dw1.5}
where  $\Delta t=0.005$. We compare the results for the simulation
in Fig.~\ref{fig:dw1.4} obtained with $\Delta t = 0.0005, 0.001$ and $0.005$.
The results for the first two values agree very well; those for
$\Delta t = 0.005$ start to differ at late times. This means that
a choice $\Delta t = 0.001$ is appropriate. In the case of Fig. 8
the variations with time are much slower, so that the
choice $\Delta t = 0.005$ is sufficient.
 
\begin{figure}[htbp]
  \centering
\hspace{-0.5cm}\includegraphics[width=7.5cm]{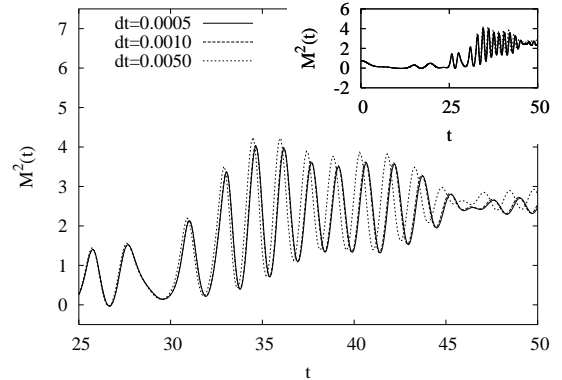}
\caption{\label{fig:time-grid}Dependence of the time evolution of
  $\calm^2(t)$ on the
  time  $\Delta t$ for the simulation with the parameters from
  Fig.~\ref{fig:dw1.4} in the time range
  $t\in[25,50]$; in the inset the whole time range is shown; 
the momentum distance is fixed at $\Delta p=0.05$ and the momentum cutoff
$p_\mathrm{max}=10$;
the solid line represents the simulation for $\Delta t=0.0005$, 
the long dashed line $\Delta t=0.001$ and the short dashed line 
$\Delta t=0.005$.
 }
\end{figure}

\end{itemize}
 We would finally like to point out that it is in no way inherent in our
numerical approach to use an equidistant momentum grid. Indeed
it is more economical to choose $\Delta p$ small for small momenta
and to let it increase for larger ones, as was done, .e.g., in
previous computations of our group. In one space dimension the
choice of equidistant momenta turns momentum conservation 
in a trivial algebra of indices. In three space dimensions one may
use the O$(3)$ invariance of the mode functions as functions of
${\bf p}$. Then, due to angular integrations, the momentum integrals
become convolutions of mode functions with phase space functions
and an equidistant momentum grid does not lead to any major 
simplification.


\begin{thebibliography}{99}

\bibitem{Berges:2001fi}
J.~Berges,
Nucl.\ Phys.\ A {\bf 699}, 847 (2002)
[arXiv:hep-ph/0105311].

\bibitem{Aarts:2002dj}
G.~Aarts, D.~Ahrensmeier, R.~Baier, J.~Berges and J.~Serreau,
Phys.\ Rev.\ D {\bf 66}, 045008 (2002)
[arXiv:hep-ph/0201308].

\bibitem{Cornwall:1974vz}
J.~M.~Cornwall, R.~Jackiw and E.~Tomboulis,
Phys.\ Rev.\ D {\bf 10}, 2428 (1974).

\bibitem{Calzetta:1987ey}
E.~Calzetta and B.~L.~Hu,
Phys.\ Rev.\ D {\bf 35}, 495 (1987);
Phys.\ Rev.\ D {\bf 37}, 2878 (1988).

\bibitem{skform}
J. S. Schwinger, J. Math. Phys. {\bf 2}, 407 (1961);
L. V. Keldysh, Zh. Eksp. Teor. Fiz. {\bf 47}, 1515 (1964)
[ Sov. Phys. -- JETP {\bf 20}, 1018 (1965)]; see also 
G. Zhou, Z. Su, B. Hao, and L. Yu,
Phys. Rep. {\bf 118}, 1 (1985).

\bibitem{Mihaila:2001sr}
B.~Mihaila, F.~Cooper and J.~F.~Dawson,
Phys.\ Rev.\ D {\bf 63}, 096003 (2001)
[arXiv:hep-ph/0006254].

\bibitem{Aarts:2001yn}
G.~Aarts and J.~Berges,
Phys.\ Rev.\ Lett.\  {\bf 88}, 041603 (2002)
[arXiv:hep-ph/0107129].

\bibitem{Blagoev:2001ze}
K.~Blagoev, F.~Cooper, J.~Dawson and B.~Mihaila,
Phys.\ Rev.\ D {\bf 64}, 125003 (2001)
[arXiv:hep-ph/0106195].


\bibitem{Cooper:2002ze}
F.~Cooper, J.~F.~Dawson and B.~Mihaila,
arXiv:hep-ph/0207346.


\bibitem{Berges:2000ur}
J.~Berges and J.~Cox,
Phys.\ Lett.\ B {\bf 517}, 369 (2001)
[arXiv:hep-ph/0006160].

\bibitem{Cooper:2002qd}
F.~Cooper, J.~F.~Dawson and B.~Mihaila,
arXiv:hep-ph/0209051.

\bibitem{Berges:2002cz}
J.~Berges and J.~Serreau,
arXiv:hep-ph/0208070.


\bibitem{Baacke:2001zt}
J.~Baacke and S.~Michalski,
Phys.\ Rev.\ D {\bf 65}, 065019 (2002)
[arXiv:hep-ph/0109137].

\bibitem{Verschelde:bs}
H.~Verschelde and M.~Coppens,
Phys.\ Lett.\ B {\bf 287}, 133 (1992).

\bibitem{Coppens:zc}
M.~Coppens and H.~Verschelde,
Z.\ Phys.\ C {\bf 58}, 319 (1993).

\bibitem{vanHees:2001ik}
H.~van Hees and J.~Knoll,
Phys.\ Rev.\ D {\bf 65}, 025010 (2002)
[arXiv:hep-ph/0107200];
Phys.\ Rev.\ D {\bf 65}, 105005 (2002)
[arXiv:hep-ph/0111193];
Phys.\ Rev.\ D {\bf 66}, 025028 (2002)
[arXiv:hep-ph/0203008].
\bibitem{Verschelde:2000ta}
H.~Verschelde and J.~De Pessemier,
Eur.\ Phys.\ J.\ C {\bf 22}, 771 (2002)
[arXiv:hep-th/0009241].

\bibitem{Verschelde:2000dz}
H.~Verschelde,
Phys.\ Lett.\ B {\bf 497}, 165 (2001)
[arXiv:hep-th/0009123].

\bibitem{Smet:2001un}
G.~Smet, T.~Vanzielighem, K.~Van Acoleyen and H.~Verschelde,
Phys.\ Rev.\ D {\bf 65}, 045015 (2002)
[arXiv:hep-th/0108163].

\bibitem{Baacke:2002pi}
J.~Baacke and S.~Michalski,
arXiv:hep-ph/0210060.

\bibitem{Okopinska:1995mi}
A.~Okopinska,
Annals Phys.\  {\bf 249}, 367 (1996)
[arXiv:hep-th/9512113].


\bibitem{Dudal:2002zn}
D.~Dudal and H.~Verschelde,
arXiv:hep-th/0210098.


\bibitem{ItzyksonZuber}
see, e.g., C. Itzykson and  J.-B. Zuber, 
{\em Quantum field theory}, McGraw-Hill Inc., New York 1980, section 1.2.2.


\bibitem{Cooper:1987pt}
F.~Cooper and E.~Mottola,
Phys.\ Rev.\ D {\bf 36}, 3114 (1987).

\bibitem{Maslov:1998bf}
V.~P.~Maslov and O.~Y.~Shvedov,
Theor.\ Math.\ Phys.\  {\bf 114}, 184 (1998)
[Teor.\ Mat.\ Fiz.\  {\bf 114}, 233 (1998)]
[arXiv:hep-th/9709151].

\bibitem{Baacke:1998zz}
J.~Baacke, K.~Heitmann and C.~Patzold,
Phys.\ Rev.\ D {\bf 57}, 6398 (1998)
[arXiv:hep-th/9711144].

\bibitem{Destri:1999hd}
C.~Destri and E.~Manfredini,
Phys.\ Rev.\ D {\bf 62}, 025007 (2000)
[arXiv:hep-ph/0001177].


\bibitem{Traschen:1990sw}
J.~H.~Traschen and R.~H.~Brandenberger,
Phys.\ Rev.\ D {\bf 42}, 2491 (1990).

\bibitem{Kofman:1994rk}
L.~Kofman, A.~D.~Linde and A.~A.~Starobinsky,
Phys.\ Rev.\ Lett.\  {\bf 73}, 3195 (1994)
[arXiv:hep-th/9405187].

\bibitem{Boyanovsky:1996sq}
D.~Boyanovsky, H.~J.~de Vega, R.~Holman and J.~F.~Salgado,
Phys.\ Rev.\ D {\bf 54}, 7570 (1996)
[arXiv:hep-ph/9608205].


\bibitem{Boyanovsky:1998zg}
D.~Boyanovsky, C.~Destri, H.~J.~de Vega, R.~Holman and J.~Salgado,
Phys.\ Rev.\ D {\bf 57}, 7388 (1998)
[arXiv:hep-ph/9711384].



\bibitem{Cooper:1997ii}
F.~Cooper, S.~Habib, Y.~Kluger and E.~Mottola,
Phys.\ Rev.\ D {\bf 55}, 6471 (1997)
[arXiv:hep-ph/9610345].


\bibitem{Griffiths}
see, e.g., R. B. Griffiths in {\em Phase Transitions and Critical Phenomena},
C. Domb and M.S. Green, Eds., Academic press, New York 1972,
Vol. 1, p. 7.

\bibitem{Calzetta:2002ub}
E.~A.~Calzetta and B.~L.~Hu,
arXiv:hep-ph/0205271.




\end{thebibliography}
\end{document}